\documentclass{SciPost}

\hypersetup{
    colorlinks,
    linkcolor={red!50!black},
    citecolor={blue!50!black},
    urlcolor={blue!80!black}
}

\usepackage[bitstream-charter]{mathdesign}
\DeclareSymbolFont{usualmathcal}{OMS}{cmsy}{m}{n}
\DeclareSymbolFontAlphabet{\mathcal}{usualmathcal}

\fancypagestyle{SPstyle}{
\fancyhf{}
\lhead{\colorbox{scipostblue}{\bf \color{white} ~SciPost Physics }}
\rhead{{\bf \color{scipostdeepblue} ~Submission }}

\fancyfoot[C]{\textbf{\thepage}}
}

\usepackage{bbm}
\usepackage{hyperref}
\usepackage[symbol]{footmisc}

\newcommand{\PP}{P}
\newcommand{\QQ}{Q}
\newcommand{\onlinecite}{\cite}

\begin{document}
%

\pagestyle{SPstyle}

\begin{center}{\Large \textbf{\color{scipostdeepblue}{
Precise quantum-geometric electronic properties from first principles \\
}}}\end{center}

\begin{center}\textbf{
Jos\'e Lu\'{\i}s Martins\textsuperscript{1,2$\star$},
Carlos L.\ Reis\textsuperscript{2}\footnote[4]{Carlos L.\ Reis passed away before the conclusion of this work.}
and Ivo Souza\textsuperscript{3,4$\dagger$}
}\end{center}

\begin{center}
{\bf 1} Departamento de F\'{\i}sica, Instituto Superior T{\'e}cnico,
Universidade de Lisboa, 1049-001 Lisboa, Portugal
\\
{\bf 2} INESC MN, 1000-029 Lisboa, Portugal
\\
{\bf 3} Centro de Física de Materiales, Universidad del País Vasco,
 20018 San Sebastían, Spain
\\
{\bf 4} Ikerbasque Foundation, 48013 Bilbao, Spain
\\[\baselineskip]
$\star$ \href{mailto:email1}{\small jlmartins@inesc-mn.pt}\,,\quad
$\dagger$ \href{mailto:email2}{\small ivo.souza@ehu.es}
\end{center}

%

\section*{\color{scipostdeepblue}{Abstract}}
\textbf{\boldmath{ The calculation of quantum-geometric properties of
    Bloch electrons~-- Berry curvature, quantum metric, orbital
    magnetic moment and effective mass~-- was implemented in a
    pseudopotential plane-wave code.  The starting point was the first
    derivative of the periodic part of the wavefunction
    $\psi_{\bf k}({\bf r})$ with respect to wavevector ${\bf k}$.
    This was evaluated with perturbation theory by solving a
    Sternheimer equation. Comparison of effective masses obtained from
    perturbation theory for silicon and gallium arsenide with
    carefully-converged numerical second derivatives of band energies
    confirmed the high precision of the method.  Calculations of
    quantum-geometric quantities for gapped graphene were performed by
    adding a bespoke symmetry-breaking potential to first-principles
    graphene.  As the two bands near the opened gap are reasonably
    isolated, the results could be compared with those obtained from
    an analytical two-band model, allowing to assess the strengths and
    limitations of such widely-used models.  The final application was
    trigonal tellurium, where some quantum-geometric quantities flip
    sign with chirality.}}

\vspace{\baselineskip}

%

\noindent\textcolor{white!90!black}{%
\fbox{\parbox{0.975\linewidth}{%
\textcolor{white!40!black}{\begin{tabular}{lr}%
  \begin{minipage}{0.6\textwidth}%
    {\small Copyright attribution to authors. \newline
    This work is a submission to SciPost Physics. \newline
    License information to appear upon publication. \newline
    Publication information to appear upon publication.}
  \end{minipage} & \begin{minipage}{0.4\textwidth}
    {\small Received Date \newline Accepted Date \newline Published Date}%
  \end{minipage}
\end{tabular}}
}}
}


\vspace{10pt}
\noindent\rule{\textwidth}{1pt}
\tableofcontents
\noindent\rule{\textwidth}{1pt}
\vspace{10pt}

\section{Introduction}

Exploring the quantum-geometric and topological properties of
electrons in crystals is a very active research
topic~\cite{xiao-rmp10,vanderbilt-book18,yu2025quantumgeometryquantummaterials,
  Gao2025quantumgeometry}.  Band effective masses, which have been a
staple of semiconductor physics since the early days of solid-state
physics, are now recognized as one of the basic quantum-geometric
properties of materials, along with the Berry curvature, quantum
metric, and intrinsic orbital moment.  A characteristic of such
quantities is that they can vary by several orders of magnitude across
the Brillouin zone (BZ), with large absolute values usually associated
with small regions near band degeneracies or quasi-degeneracies.
Therefore, a very fine sampling of the BZ may be required to find
those regions, or to calculate properties that are given by BZ
integrals of expressions involving quantum-geometric
quantities~\cite{yao-prl04}. Interpolation methods, in particular
Wannier interpolation, have been extensively used to calculate such
properties~\cite{wang-prb06,lopez-prb12,marrazzo-rmp24}, although the
evaluation of the quantum metric in this manner was implemented
only recently~\cite{urru-prb25}.  In some cases~-- for example, the
band edges of semiconductors~-- the region of interest in the BZ may
be known, allowing the calculation to be done without recourse to an
interpolation scheme.

Effective masses, the inverse of the second derivatives of band
energies with respect to wavevector, can be calculated by finite
differences~\cite{Gonze_mass_2016}.  However, in many cases the
spin-orbit interaction introduces strong localized perturbations in
the band dispersions. Under those circumstances it becomes nontrivial
to find good parameters for the numerical differentiation, and to
disentangle real from avoided band
crossings~\cite{BZ_integration_Pickard_1999}.  Adjusting empirical
parameters of ${\bf k}\cdot{\bf p}$-based models is another
possibility, although it requires the prior step of setting up an
effective Hamiltonian~\cite{kdotp_nitrides_JRLeite_2001}.  Recently,
degenerate perturbation theory was used to calculate precise effective
masses from first principles with pseudopotential and linearized
augmented-plane-wave (LAPW)
codes~\cite{Gonze_mass_2016,Blaha_mass_2021}, avoiding the
difficulties with numerical differentiation or model fitting.

Here, the effective masses are viewed as just one of several related
quantum-geometric quantities that can be calculated from the first
wavevector derivatives of the cell-periodic Bloch wavefunctions.
Other quantities that are obtainable in this way include the Berry
curvature, the quantum metric, and the orbital magnetic
moment~\cite{xiao-rmp10,vanderbilt-book18}, and finite-difference
schemes for computing them are
available~\cite{marzari-prb97,ceresoli-prb06}.  While the band
effective masses and the quantum metric are generically nonzero, the
Berry curvature and the orbital magnetic moment vanish identically if
both spatial-inversion and time-reversal symmetry are
present~\cite{xiao-rmp10}. In this work, the evaluation of all these
quantities is implemented in a pseudopotential plane-wave code using
perturbation theory.  The code is open source and available for
download~\cite{cpw2000}.

The paper is organized as follows.  In Sec.~\ref{sec:method} the
context, method and implementation are presented.  Special attention
is paid to the treatment of band degeneracies, as degenerate
perturbation theory is much more complex to implement, and the
physical tensors have degeneracy indices added to the spatial indices.
To illustrate the method, the precise effective masses in
centrosymmetric silicon and noncentrosymmetric gallium arsenide are
presented in Sec.~\ref{sec:results}; the convergence of the
finite-differences and ${\bf k}\cdot{\bf p}$ methods is also
discussed, using our perturbative implementation as a reference.
Section~\ref{sec:graphene} deals with an artificial system obtained by
breaking the inversion symmetry of graphene, thereby opening a gap at
the Dirac points.  This allows for a detailed comparison of the
calculated {\it ab initio} quantum-geometric quantities with
analytical results from a two-band effective model.  The final
example, in Sec.~\ref{sec:tellurium}, is trigonal tellurium, a chiral
semiconducting crystal with striking electronic responses driven by
quantum geometry at the band
edge~\cite{Te_LAPW_Snyder2014,lin-natcomms16,Tsirkin_Te_2018,sahin-prb18,furukawa-prb21,nakazawa-prm24,pan-prb25}.
The article concludes in Sec.~\ref{sec:conclusions} with a summary.

\section{Method}
\label{sec:method}

\subsection{Periodic potential}

Once the Kohn-Sham equations~\cite{KohnSham1965} of density-functional
theory (DFT)~\cite{HohenbergKohn1964} are solved and an effective
potential $V_\text{eff}({\bf r})$
is determined within some approximation, one obtains the Schrödinger
equation for that potential,
\begin{equation}
H  \psi_{d n}({\bf r}) =
  \left( -\frac{1}{2}\nabla^2 + V_\text{eff}({\bf r}) \right) \psi_{dn}({\bf r})
                     = E_n  \psi_{d n}({\bf r})\,,
  \label{eq:Horder0}
\end{equation}
with one-electron energy levels $E_n$ and wavefunctions
$\psi_{d n}({\bf r})$.  The energies are assumed to be in strictly
increasing order, $E_1 < E_2 < E_3 < \ldots$, and each energy level
has degeneracy $D_n$, with $d = 1,2,\ldots,D_n$ an additional
wavefunction index.  Wavefunctions within a degenerate subspace are
chosen to be orthonormal:
$\langle \psi_{d n} \vert \psi_{d^\prime m} \rangle = \delta_{d
  d^\prime}\delta_{n m}$, where the normalization volume is one
crystal cell.  Here and throughout the paper, we use Hartree atomic
units ($\hbar = m_e = |e| = 4 \pi \epsilon_0 = 1$) to simplify the
notation, unless stated otherwise.

For a perfect crystal the potential $V_\text{eff}$ is periodic, and
one can use Bloch's theorem to choose solutions of the Schrödinger
equation that transform according to the representations of the
translation symmetry group, labeled by the wavevector ${\bf k}$,
\begin{equation}
   H \vert \psi_{d n {\bf k}} \rangle  = E_{n {\bf k}} \vert \psi_{d n {\bf k}} \rangle.
\end{equation}
In this section, the notation for subscripts will follow the rule of
going from the more specific to the more general.  For example, in the
above equation, the subscripts indicate the function $d$ in the
eventually degenerate energy level $n$ of wavevector ${\bf k}$. For
any given reciprocal-lattice vector ${\bf G}$, the representations
${\bf k}$ and ${\bf k} + {\bf G}$ are the same and the energy bands
are periodic in reciprocal space,
\begin{equation}
    E_{n {\bf k}} =  E_{n {\bf k+ {\bf G}}}.
\end{equation}
The Bloch wavefunction can be expressed as the product of a periodic
function $u_{d n {\bf k}}({\bf r})$ and a spatial phase factor,
\begin{equation}
    \psi_{d n {\bf k}}({\bf r})  = e^{i {\bf k} \cdot {\bf r}}  u_{d n {\bf k}}({\bf r}).
\end{equation}
If ${\bf R}$ is a translation vector of the lattice, then
$u_{d n {\bf k}}({\bf r}) = u_{d n {\bf k}}({\bf r}+{\bf R})$.  The
cell-periodic parts of the Bloch wavefunctions are also orthonormal,
$\langle u_{d n {\bf k}} \vert u_{md^\prime{\bf k}} \rangle =
\delta_{n m}\delta_{d d^\prime}$, and they satisfy the following
eigenvalue equation,
\begin{equation}
   H_{\bf k} \vert u_{d n {\bf k}} \rangle
         = \left[-\frac{1}{2}\left(\nabla^2 +
                 2i{\bf k}\cdot {\boldsymbol\nabla} - k^2  \right) +
                 V_\text{eff}\right]\vert  u_{d n {\bf k}} \rangle
         = E_{n{\bf k}} \vert u_{d n {\bf k}} \rangle.
\label{eq:Hk}
\end{equation}
In the following discussion, we may omit the ``periodic part''
qualifier when referring to $u_{d n {\bf k}}({\bf r})$ if the meaning
is clear from the context.

In the present work we make the pseudopotential approximation, which
allows the use of a plane-wave basis set at the cost of introducing a
nonlocal pseudopotential~\cite{martin-book04}. Following Kleinman and
Bylander (KB)~\cite{KleinmanBylander1982}, the nonlocal part of the
pseudopotential is represented by (${\bf k}$-dependent) ``KB
projectors'' $\vert p_{j{\bf k}} \rangle$, and the final form of the
eigenvalue equation to be solved is
\begin{equation}
    H_{\bf k} \vert u_{d n {\bf k}} \rangle
        = \left[ -\frac{1}{2}\Bigl(\nabla^2 +
              2i{\bf k}\cdot {\boldsymbol\nabla} - k^2  \Bigr)
                 + V_\text{eff}  + \sum_j \vert p_{j{\bf k}} \rangle w_j \langle p_{j{\bf k}} \vert \right]
                          \vert u_{d n {\bf k}} \rangle
        = E_{n{\bf k}} \vert u_{d n {\bf k}} \rangle.
\label{eq:kdotp}
\end{equation}
The weights $w_j$ can be positive or negative, and may be assumed to
be just $\pm 1$ if the KB projectors are not normalized.

The Bloch eigenfunctions are not unique: if
$\left\{\vert {u}_{d n {\bf k}}\rangle\right\}$ is a set of solutions
of Eq.~\eqref{eq:kdotp} for a given energy level $E_{n\bf k}$, then
the states
\begin{equation}
\vert {\tilde u}_{d n {\bf k}} \rangle =
\sum_{d^\prime = 1}^{D_{n {\bf k}}}   \vert {u}_{d^\prime n {\bf k}} \rangle U_{d^\prime d}(n,{\bf k}) ,
   \label{eq:unittransf}
\end{equation}
with $U_{d^\prime d}(n,{\bf k})$ an arbitrary unitary matrix, are also
solutions of Eq.~\eqref{eq:kdotp} satisfying the orthonormality
constraints.  This gauge arbitrariness has to be acknowledged when
considering the ${\bf k}$ dependence of the wavefunctions.

\subsection{Perturbation theory}

Approximations for energies and wavefunctions in the neighborhood of a
${\bf k}$ point can be obtained from Eq.~\eqref{eq:kdotp} using
perturbation theory.  The main ingredients are the derivatives of the
Hamiltonian $H_{\bf k}$ with respect to the Cartesian components
$k_\alpha$ of the ${\bf k}$ vector.  We will use Greek letters to
indicate Cartesian directions, and place Cartesian indices as
subscripts or superscripts in accordance with Einstein convention.
The first derivative of the Hamiltonian reads
\begin{equation}
H_{\bf k}^\alpha \equiv \partial_\alpha H_{\bf k}
                    = -i \frac{\partial}{\partial r^\alpha} + k_\alpha
                              + \Bigl(\sum_j \vert p_{j{\bf k}}^\alpha \rangle w_j \langle p_{j{\bf k}} \vert
                                    + \sum_j \vert p_{j{\bf k}} \rangle w_j \langle p_{j{\bf k}}^\alpha \vert\Bigr),
   \label{eq:Horder1}
\end{equation}
where $\partial_\alpha = {\partial}/{\partial k_\alpha}$ and
$\vert p_{j{\bf k}}^\alpha \rangle = \partial_\alpha \vert p_{j{\bf
    k}} \rangle$.  The second derivative is\footnote[4]{The
  derivatives of the pseudopotential projectors are calculated in the
  code by just applying the chain rule, with some care on how
  intermediate quantities are defined to ensure easy parallelization.
  The code uses lattice coordinates, and therefore a metric matrix
  appears in the expressions; for example, the Kronecker delta in
  Eq.~\eqref{eq:Horder2} becomes a metric tensor.  Any apparent
  inconsistencies with the Einstein notation in equations in this
  paper are indicative of the presence of the metric in the code.}
\begin{equation}
   \begin{aligned}
   H_{\bf k}^{\alpha\beta} \equiv \partial_\alpha \partial_\beta H_{\bf k}
                              = \delta^{\alpha \beta}
                                        &+ \Bigl(\sum_j \vert p_{j{\bf k}}^{\alpha\beta} \rangle w_j \langle p_{j{\bf k}} \vert
                                             + \sum_j \vert p_{j{\bf k}}^{\alpha} \rangle w_j \langle p_{j{\bf k}}^{\beta} \vert \\
                                             &
                                             \;\;\,
                                             + \sum_j \vert p_{j{\bf k}} \rangle w_j \langle p_{j{\bf k}}^{\alpha\beta} \vert
                                             + \sum_j \vert p_{j{\bf k}}^{\beta} \rangle w_j \langle p_{j{\bf k}}^{\alpha} \vert  \Bigr),
   \end{aligned}
   \label{eq:Horder2}
\end{equation}
where
$\vert p_{j{\bf k}}^{\alpha\beta} \rangle
=\partial_\alpha\partial_\beta \vert p_{j{\bf k}} \rangle$.

The first-order equation of degenerate perturbation theory
is~\cite{vanderbilt-book18,MIT-OCW_Zwiebach_2018}
\begin{equation}
   ({E}_{n{\bf k}} - H_{\bf k}) \vert {u}_{d n {\bf k}}^\alpha \rangle =
         \bigl( H_{\bf k}^\alpha - E_{d n {\bf k}}^\alpha \bigr)
                 \vert {u}_{d n {\bf k}} \rangle,
   \label{eq:1storder}
\end{equation}
with $E_{d n {\bf k}}^\alpha$ and
$\vert {u}_{d n {\bf k}}^\alpha \rangle$ the first-order corrections
to the energies and wavefunctions, respectively.  Multiplying
Eq.~\eqref{eq:1storder} from the left by
$\langle {u}_{d^\prime n {\bf k}}\vert$ leads to the condition
\begin{equation}
0 = \langle {u}_{d^\prime n {\bf k}}| H_{\bf k}^\alpha \vert {u}_{d n {\bf k}} \rangle
- E_{d n {\bf k}}^\alpha \delta_{d d^\prime}.
\label{eq:diagpert}
\end{equation}
The first-order corrections to the band energies are therefore
obtained by diagonalizing the matrix of the perturbation in the
degenerate subspace of dimension $D_{n {\bf k}}$.

The operator ${E}_{n{\bf k}} - H_{\bf k}$ has a null space, and the
components of $\vert {u}_{d n {\bf k}}^\alpha \rangle$ on that
subspace cannot be determined at this level of perturbation
theory~\cite{MIT-OCW_Zwiebach_2018}.  However, those components are
not needed to calculate the quantum-geometric quantities to be
introduced in Sec.~\ref{sec:quantum-geom}.  For that purpose, it is
convenient to define for each energy level $E_{n{\bf k}}$ the
projector onto its eigenspace, which is the null space of
${E}_{n{\bf k}} - H_{\bf k}$, as well as the projector onto the
complement space,
\begin{equation}
   \PP_{n {\bf k}} = \sum_{d=1}^{D_{n {\bf k}}} \vert {u}_{d n {\bf k}} \rangle \langle {u}_{d n {\bf k}} \vert,
       \qquad
   \QQ_{n {\bf k}} = {\mathbbm 1} -  \PP_{n {\bf k}}.
\end{equation}
With these projectors in hand, it is now possible to replace
Eq.~\eqref{eq:1storder} for $\vert {u}_{d n {\bf k}}^\alpha \rangle$
with a Sternheimer equation for
$\QQ_{n {\bf k}} \vert {u}_{d n {\bf
    k}}^\alpha\rangle$~\cite{vanderbilt-book18},
\begin{equation}
   ({E}_{n {\bf k}} - H_{\bf k}) \QQ_{n {\bf k}} \vert {u}_{d n {\bf k}}^\alpha \rangle =
          \QQ_{n {\bf k}} H_{\bf k}^\alpha \vert {u}_{d n {\bf k}} \rangle.
   \label{eq:stern2}
\end{equation}
Thanks to the disappearance of the highly nonlinear first-order energy
term $E_{d n {\bf k}}^\alpha$ from the right-hand side, this is now a
linear equation; therefore, any linear combination of the right-hand
side will have as a solution the same linear combination of
wavefunction perturbations.  This means that once solutions have been
found for three linearly-independent directions, the solution for any
direction can be recovered.

With the wavefunctions expanded in a basis set, Eq.~\eqref{eq:stern2}
is just a linear system, which can in principle be handled by any
linear-algebra package.  However, when using plane waves the basis
sets tend to be very large.  Nevertheless, the calculation of
$H_{\bf k} \vert u \rangle$ for an arbitrary wavefunction
$\vert u \rangle$ can still be done very
efficiently~\cite{CarParrinello1985,MartinsCohen1988}. This suggests
an iterative solution method for Eq.~\eqref{eq:stern2}, requiring an
appropriate initial guess and a stable iterative procedure.  Since
that equation has the known closed-form
solution~\cite{vanderbilt-book18}
\begin{equation}
   \QQ_{n {\bf k}} \vert {u}_{d n {\bf k}}^\alpha \rangle = \sum_{m \neq n} \sum_{d^\prime=1}^{D_{m {\bf k}}}
               \frac{\vert {u}_{m d^\prime{\bf k}} \rangle \langle {u}_{m d^\prime{\bf k}} \vert}
                    {{E}_{m{\bf k}} - {E}_{n{\bf k}}}
                       H_{\bf k}^\alpha \vert {u}_{d n {\bf k}} \rangle,
   \label{eq:sternsol}
\end{equation}
a truncation of the sum over the energy levels $m$ provides a good
starting point for the iterative procedure.  A conjugate gradient
method that only requires the computation of
$H_{\bf k} \vert u \rangle$ for an arbitrary $\vert u \rangle$ will be
computationally efficient.

The conjugate gradient method for linear systems is only guaranteed to
be stable if the matrix is positive- or negative-definite, which is
not the case for Eq.~\eqref{eq:stern2} except when $n$ is the lowest
energy level.  Fortunately, it is easy to recast the equation in a
form that is stable, as follows~\cite{Gonze_mass_2016}:
define, for fixed $n,d,\alpha$ and a cutoff
$N \geq n$, a decomposition of the desired wavefunction derivatives
into an inner space and an outer space,
\begin{equation}
   \QQ_{n {\bf k}} \vert {u}_{d n {\bf k}}^\alpha \rangle = \vert {u}_{d n {\bf k}}^\alpha \rangle_\text{in} + \vert {u}_{d n {\bf k}}^\alpha \rangle_\text{out}.
\end{equation}
In the inner space, we use a truncation of the closed-form solution of
Eq.~\eqref{eq:sternsol},
\begin{equation}
   \vert {u}_{d n {\bf k}}^\alpha \rangle_\text{in} = \sum_{\substack{m=1 \\ m \neq n}}^N \sum_{d=1}^{d_m}
               \frac{\vert {u}_{m d{\bf k}} \rangle \langle {u}_{m d{\bf k}} \vert}{{E}_{n{\bf k}} - {E}_{m{\bf k}}}
                     H_{\bf k}^\alpha  \vert u_{d n {\bf k}} \rangle.
\end{equation}
Defining the ${\bf k}$-dependent projectors
\begin{equation}
   \PP_\text{in} = \sum_{n=1}^N \sum_{d=1}^{D_{n {\bf k}}} \vert {u}_{d n {\bf k}} \rangle \langle {u}_{d n {\bf k}} \vert,
       \qquad
   \QQ_\text{in} = {\mathbbm 1} -  \PP_\text{in} = \PP_\text{out},
\end{equation}
one obtains a Sternheimer equation for
$\vert {u}_{d n {\bf k}}^\alpha \rangle_\text{out} = \QQ_\text{in}\vert
    {u}_{d n {\bf k}}^\alpha \rangle$,
\begin{equation}
   ({E}_{n{\bf k}} - H_{\bf k})\QQ_\text{in} \vert {u}_{d n {\bf k}}^\alpha \rangle =
          \QQ_\text{in} H_{\bf k}^\alpha \vert {u}_{d n {\bf k}} \rangle.
\end{equation}
Since for $n \leq N$ the operator on the left-hand side is represented
by a negative-definite matrix, the above linear equation is stable
with a conjugate gradient solver.  In practice, $N$ is chosen as the
highest band for which the quantum-geometric quantities of interest
are to be calculated.

\subsection{Band velocities}

For each energy level ${E}_{n {\bf k}}$, one can build at first order
in perturbation theory the following ``vector of matrices,'' or tensor
with three indices, which already appeared in Eq.~\eqref{eq:diagpert},
\begin{equation}
  {\epsilon}_{d d^\prime}^\alpha (n,{\bf k})
                      = \langle {u}_{d n {\bf k}} \vert H_{\bf k}^\alpha \vert {u}_{d^\prime n {\bf k}} \rangle,
              \qquad d,d^\prime = 1,\ldots,D_{n {\bf k}}.
\end{equation}
In the nondegenerate case ($D_{n {\bf k}} = 1$), this immediately
gives the first derivative of the energy with respect to $k_\alpha$,
i.e., the band velocity
\begin{equation}
  {E}_{n {\bf k}}^\alpha = {\partial_\alpha} {E}_{n {\bf k}}
                      = \langle {u}_{n{\bf k}} \vert H_{\bf k}^\alpha \vert {u}_{n{\bf k}} \rangle,
\end{equation}
where the redundant $d$ index was omitted.

When $D_{n {\bf k}}>1$ for a given energy level $E_{n {\bf k}}$, it is
necessary to first pick a direction $\hat {\bf q}$ in ${\bf k}$ space
(with $|\hat {\bf q}| = 1$), and then construct the first-order
perturbation matrix
\begin{equation}
  A_{d d^\prime}^{(1)}(\hat {\bf q},n,{\bf k}) = \sum_\alpha {\hat q}_\alpha {\epsilon}_{d d^\prime}^\alpha(n,{\bf k}),
  \label{eq:firstorderE}
\end{equation}
whose eigenvalues $E_{j n{\bf k}}^{(1)}(\hat{\bf q})$ are the band
velocities along that particular direction.
(Equation~\eqref{eq:firstorderE} is the matrix representation of the
first-order Hamiltonian
$H_{\bf k}^{\hat{\bf q}} = \sum_\alpha \hat q_\alpha H_{\bf k}^\alpha$
within the degenerate subspace.)

\subsection{Quantum-geometric quantities}
\label{sec:quantum-geom}

Once we have $H_{\bf k}$, $H_{\bf k}^{\alpha }$,
$H_{\bf k}^{\alpha\beta}$, and
$\QQ_{n {\bf k}} \vert {u}_{d n {\bf k}}^\alpha \rangle$, we can
calculate several quantum-geometric quantities, including effective
masses, with just algebraic operations. Given a group of $D_{n\bf k}$
degenerate states at ${\bf k}$, one can associate
with it a quantum-geometric tensor defined
as~\cite{vanderbilt-book18,yu2025quantumgeometryquantummaterials,marzari-prb97}
\begin{equation}
T_{d d^\prime}^{\alpha\beta}(n,{\bf k}) = \langle {u}_{d n {\bf k}}^\alpha
\vert \QQ_{n {\bf k}} \vert {u}_{d^\prime n {\bf k}}^\beta \rangle,
\label{eq:T}
\end{equation}
which transforms covariantly under the gauge transformation of
Eq.~\eqref{eq:unittransf}.  This tensor is Hermitian with respect to
the interchange of pairs of indices
$(d,\alpha) \leftrightarrow (d^\prime,\beta)$.  It can be split into
symmetric and antisymmetric parts in the Cartesian indices, or
equivalently, into Hermitian and anti-Hermitian parts in the
degeneracy indices.  The non-Abelian Berry curvature is
\begin{equation}
   \Omega_{d d^\prime}^{\alpha\beta}(n,{\bf k}) = i \, T_{d d^\prime}^{\alpha\beta}(n,{\bf k})
                                                 -i \, T_{d d^\prime}^{\beta\alpha}(n,{\bf k})
                                                = i \, T_{d d^\prime}^{\alpha\beta}(n,{\bf k})
                                                 -i \, \bigl(T_{d^\prime d}^{\alpha\beta}(n,{\bf k})\bigr)^*,
   \label{eq:curvature}
\end{equation}
and the non-Abelian quantum metric is
\begin{equation}
   g_{d d^\prime}^{\alpha\beta}(n,{\bf k}) = \frac{1}{2} T_{d d^\prime}^{\alpha\beta}(n,{\bf k})
                                             + \frac{1}{2} T_{d d^\prime}^{\beta\alpha}(n,{\bf k})
                                           = \frac{1}{2} T_{d d^\prime}^{\alpha\beta}(n,{\bf k})
                                             + \frac{1}{2} \bigl(T_{d^\prime d}^{\alpha\beta}(n,{\bf k})\bigr)^*.
   \label{eq:metric}
\end{equation}
With these definitions we have
\begin{equation}
   T_{d d^\prime}^{\alpha\beta}(n,{\bf k}) = g_{d d^\prime}^{\alpha\beta}(n,{\bf k})
                                             - i \frac{1}{2} \Omega_{d d^\prime}^{\alpha\beta}(n,{\bf k}).
\end{equation}
For a nondegenerate level, the quantum metric and the Berry curvature
become real-symmetric and real-antisymmetric Cartesian tensors,
respectively~\cite{ProvostVallee_1980,berry1989},
\begin{equation}
   g^{\alpha\beta}(n,{\bf k}) = \text{Re} \langle {u}_{n {\bf k}}^\alpha \vert {u}_{n {\bf k}}^\beta \rangle
                            - \langle {u}_{n {\bf k}}^\alpha \vert {u}_{n {\bf k}} \rangle
                              \langle {u}_{n {\bf k}} \vert {u}_{n {\bf k}}^\beta \rangle .
\end{equation}
and
\begin{equation}
   \Omega^{\alpha\beta}(n,{\bf k}) = -2 \text{Im} \langle {u}_{n{\bf k}}^\alpha  \vert {u}_{n{\bf k}}^\beta \rangle .
\end{equation}
Moreover, the latter can be transformed into a
pseudovector as
$\Omega^\gamma(n,{\bf k}) = \varepsilon_{\alpha \beta \gamma}
\Omega^{\alpha\beta}(n,{\bf k})$.

Another quantity of interest is the ``mass-moment tensor''
\begin{equation}
\Gamma_{d d^\prime}^{\alpha\beta}(n,{\bf k}) = \langle {u}_{d n {\bf k} }^\alpha \vert \QQ_{n {\bf k}}
(H_{\bf k}-E_{n {\bf k}}) \QQ_{n {\bf k}} \vert {u}_{d^\prime n {\bf k}}^\beta \rangle,
\label{eq:Gamma}
\end{equation}
whose trace over orbital indices (spanning the full set of occupied
states, not just a degenerate group) appears in a generalized
oscillator-strength sum rule with both time-even and time-odd
parts~\cite{souza-prb08}.  This quantity has the same symmetry
properties as $T_{d d^\prime}^{\alpha\beta}$, and can be partitioned
along similar lines.  The Cartesian antisymmetric (or band
anti-Hermitian) part is the intrinsic orbital magnetic moment tensor
of the degenerate group of bands~\cite{chang-jpcm08,xiao-rmp10},
\begin{equation}
   {\mathfrak m}_{d d^\prime}^{\alpha\beta}(n,{\bf k}) = \frac{|e|}{\hbar}
         \Bigl(\frac{1}{2i} \Gamma_{d d^\prime}^{\alpha\beta}(n,{\bf k})
                 -\frac{1}{2i} \Gamma_{d d^\prime}^{\beta\alpha}(n,{\bf k})\Bigr),
  \label{eq:orbmagmom}
\end{equation}
while the Cartesian symmetric (or band Hermitian) part contributes to
a generalized inverse effective mass tensor of the band
group, which can be conveniently written as follows,
\begin{equation}
    {\epsilon}_{d d^\prime}^{\alpha\beta}(n,{\bf k}) = \hbar^2\Bigl(\frac{1}{m^*}\Bigr)_{d d^\prime}^{\alpha\beta}(n,{\bf k})
            = \langle {u}_{d n {\bf k}} \vert H_{\bf k}^{\alpha\beta} \vert {u}_{d^\prime n {\bf k}} \rangle - \bigl(\Gamma_{d d^\prime}^{\alpha\beta}(n,{\bf k}) + \Gamma_{d d^\prime}^{\beta\alpha}(n,{\bf k}) \bigr).
  \label{eq:invmass}
\end{equation}
(Here we restored the constants $\hbar$ and $|e|$ for clarity.)  For a
nondegenerate level, the orbital magnetic moment and inverse
effective mass tensors
become~\cite{chang-jpcm08,xiao-rmp10,Resta_2018}
\begin{equation}
{\mathfrak m}^{\alpha\beta}(n,{\bf k}) = - \frac{|e|}{\hbar}\text{Im} \langle {u}_{n {\bf k}}^\alpha \vert
H_{\bf k} -{E}_{n {\bf k}}
\vert {u}_{n {\bf k}}^\beta \rangle
\label{eq:orbmagmomnondeg}
\end{equation}
and
\begin{equation}
E^{\alpha\beta}_{n\bf k}=    \partial_\alpha \partial_\beta E_{n {\bf k}} =
\hbar^2\Bigl(\frac{1}{m^*}\Bigr)^{\alpha\beta}(n,{\bf k})
            = \langle {u}_{n {\bf k}} \vert H_{\bf k}^{\alpha\beta} \vert {u}_{n {\bf k}} \rangle -
            2 \text{Re}\langle {u}_{n {\bf k} }^\alpha \vert
                             H_{\bf k}-E_{n {\bf k}} \vert {u}_{n {\bf k}}^\beta \rangle.
  \label{eq:massnondeg}
\end{equation}
Inserting Eq.~\eqref{eq:sternsol} in Eq.~\eqref{eq:massnondeg}, and
specializing to a local potential, yields the familiar sum-over-states
expression for the inverse effective mass tensor.

\begin{figure} 
\centering
\includegraphics[width=1.0\columnwidth]{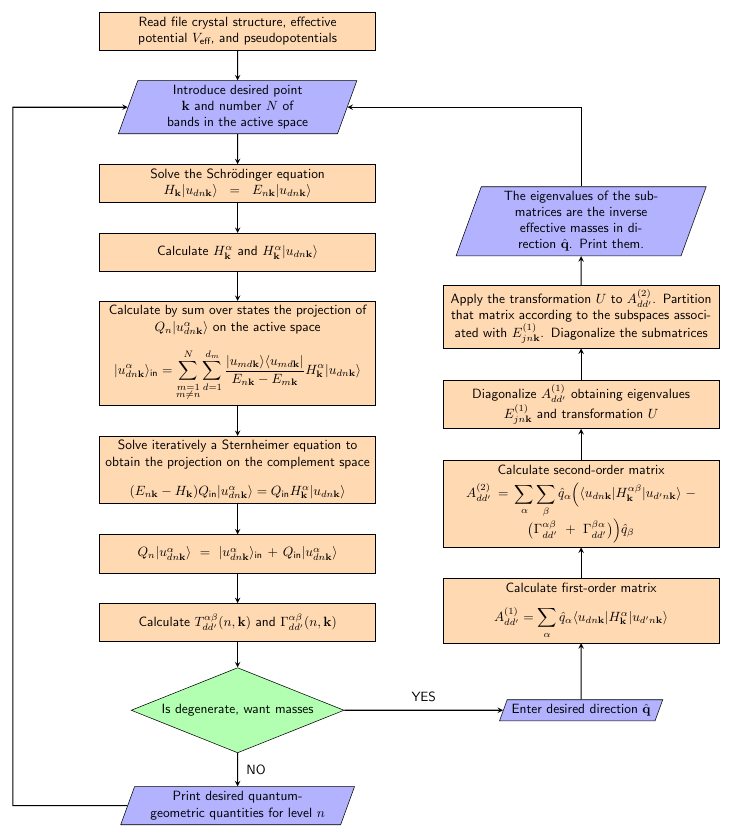}
\caption{A flowchart of the procedure for calculating the
  quantum-geometric quantities and transport equivalent effective
  masses.  The draft user guide, with the instructions to perform
  these calculations and interpret the results, is available from the
  same website as the code~\cite{cpw2000}.}
\label{fig:flowchart}
\end{figure}

\subsection{Calculation of effective masses in a given direction}
\label{subsec:massdirection}

In the absence of degeneracies, the calculation of effective masses is
straightforward, as in this case the inverse effective mass is just a
rank-two Cartesian tensor, Eq.~\eqref{eq:massnondeg}.  For degenerate
levels, the inverse effective mass is a more complicated tensor,
Eq.~\eqref{eq:invmass}, with two indices on the Cartesian coordinates
and two additional degeneracy indices.  One way to deal with this
tensor is to calculate the effective masses for a given direction
$\hat{\bf q}$ and energy level $E_{n \bf k}$; these are known as the
transport equivalent effective
masses~\cite{Gonze_mass_2016,Blaha_mass_2021,Band_warp_Resca2014}.

The starting ingredients are the matrix of first derivatives
$A_{d d^\prime}^{(1)}(\hat {\bf q},n,{\bf k})$ defined in
Eq.~\eqref{eq:firstorderE}, and the corresponding matrix of second
derivatives:
\begin{equation}
   A_{d d^\prime}^{(2)}(\hat {\bf q},n,{\bf k}) = \sum_{\alpha,\beta}
            {\hat q}_\alpha  {\epsilon}_{d d^\prime}^{\alpha\beta}(n,{\bf k})  {\hat q}_\beta,
   \label{eq:A2dd}
\end{equation}
with ${\epsilon}_{d d^\prime}^{\alpha\beta}(n,{\bf k})$ taken from
Eq.~\eqref{eq:invmass}.  To proceed, we must find the unitary
transformation $U_{d d^\prime}(\hat {\bf q},n,{\bf k})$ that
diagonalizes $A_{d d^\prime}^{(1)}(\hat {\bf q},n,{\bf k})$, written
concisely as
\begin{equation}
    U^\dagger A^{(1)} U = E^{(1)}.
\end{equation}
Here, $E_{n {\bf k}}^{(1)}(\hat{\bf q})$ is a diagonal matrix with
ordered elements
$E_{1 n {\bf k}}^{(1)}(\hat{\bf q}) < E_{2 n {\bf k}}^{(1)}(\hat{\bf
  q}) < \ldots < E_{J_{n {\bf k}} n {\bf k}}^{(1)}(\hat{\bf q})$.
There are $J_{n {\bf k}}(\hat{\bf q})$ distinct eigenvalues of
$A^{(1)}(\hat{\bf q} ,n, {\bf k})$, each with degeneracy
$I_{j n {\bf k}}(\hat{\bf q})$, satisfying
$\sum_{j = 1}^{J_{n {\bf k}}(\hat{\bf q})} I_{j n {\bf k}}(\hat{\bf
  q}) = D_{n {\bf k}}$.  The same transformation must now be applied
to the second-order matrix
$A_{d d^\prime}^{(2)}(\hat {\bf q},n,{\bf k})$,
\begin{equation}
U^\dagger A^{(2)} U = {\tilde A}^{(2)},
\end{equation}
yielding a new second-order matrix
${\tilde A}_{d d^\prime}^{(2)}(\hat {\bf q},n,{\bf k})$. Finally, for
each band velocity $E_{j n{\bf k}}^{(1)}(\hat{\bf q})$
we extract from the
associated rows and columns of
${\tilde A}_{d d^\prime}^{(2)}(\hat {\bf q},n,{\bf k})$ the submatrix
${\tilde{\tilde A}}_{d d^\prime}^{(2)}(\hat {\bf q},j, n,{\bf k})$ of
dimensions $I_{j n {\bf k}} \times I_{j n {\bf k}}$.  The eigenvalues
$E_{i j n{\bf k}}^{(2)}(\hat{\bf q})$ of
${\tilde{\tilde A}}_{d d^\prime}^{(2)}(\hat {\bf q},j, n,{\bf k})$ are
the inverse effective masses along $\hat{\bf q}$,
\begin{equation}
     \left. \frac{\partial^2}{\partial \eta^2} E_{n,{\bf k}
             + \eta \hat {\bf q}}(i,j)\right|_{\eta=0}
                     = E_{i j n{\bf k}}^{(2)}(\hat{\bf q}).
   \label{eq:secondderiv}
\end{equation}
A schematic flowchart of the procedure is shown in
Fig.~\ref{fig:flowchart}.

We note that the procedure outlined above is equivalent to applying
the unitary transformation $U$ that diagonalizes $A^{(1)}$ to the
original wavefunctions $\vert {u}_{j n{\bf k}}(\hat {\bf q}) \rangle$
to obtain a new set of wavefunctions
$\vert {\tilde u}_{i j n{\bf k}}(\hat {\bf q}) \rangle$, with
$i = 1,\ldots,I_{j n {\bf k}}(\hat{\bf q})$ and
$j=1,\ldots,J_{n\bf k}(\hat{\bf q})$; the new set is then used to
recalculate Eqs.~\eqref{eq:Gamma} and~\eqref{eq:invmass}, as they are
the relevant wavefunctions for carrying out second-order degenerate
perturbation theory~\cite{MIT-OCW_Zwiebach_2018} in the subspace with
first-order eigenvalue corrections
$E_{j n{\bf k}}^{(1)}(\hat{\bf q},n,{\bf k})$.  However, applying the
transformations directly to the operators is computationally more
efficient.

Effective masses of degenerate bands have been calculated from first
principles using perturbative methods (as opposed to finite
differences) with the {\sc abinit}~\cite{Gonze_mass_2016} and
WIEN2k~\cite{Blaha_mass_2021} codes; for the nondegenerate case, there
is earlier work~\cite{kdotp_masses_Pickard2000}.  As those
calculations, as well as the present work, use perturbation theory to
calculate effective masses, and their basis sets are all related to
plane waves -- linearized augmented-plane-wave (LAPW), projected
augmented wave (PAW) or just elementary plane waves -- there are close
similarities between all three implementations.

The main distinction between the present approach and previous works
is that we use
$\QQ_{n {\bf k}} \vert {u}_{d n {\bf k}}^\alpha \rangle$ given by the
Sternheimer equation~\eqref{eq:stern2} as our basic ingredient,
whereas the previous works start from explicit sums over states; the
connection between the two approaches is Eq.~\eqref{eq:sternsol}.  The
{\sc abinit} implementation evaluates the effective masses in a
plane-wave basis via the expression in Eq.~(66) of
Ref.~\onlinecite{Gonze_mass_2016}; that expression is different from
our Eq.~\eqref{eq:invmass} with
$\Gamma_{d d^\prime}^{\alpha\beta}(n,{\bf k})$ defined by
Eq.~\eqref{eq:Gamma}, although it is possible to show they are
equivalent using Eq.~\eqref{eq:sternsol}.  As the {\sc abinit}
implementation uses a Sternheimer equation to circunvent an infinite
sum over states, there are also points in common at the algorithmic
level.  The present approach has the advantage of yielding the Berry
curvature, quantum metric and orbital moment for an almost negligible
extra computational cost, presenting the calculation of
quantum-geometric quantities in a unified framework.

\section{Semiconductor effective masses}
\label{sec:results}

\subsection{Silicon}

As a first example, we study the effective masses of Si, with special
emphasis on the top of the valence band at $\Gamma$
(${\bf k} = {\bf 0}$).  Without spin-orbit coupling and ignoring spin,
the top of the valence band consists of three degenerate states with
$p$ character around each of the two atoms in the primitive cell.
With spin-orbit interaction, spin cannot be ignored, and the six
states at $\Gamma$ split into a twice-degenerate level with lower
energy (the ``split-off hole'' band), and four degenerate states which
become the true valence-band maximum.  Although the spin-orbit
splitting at $\Gamma$ is small, it is of the order of the energy
associated with room temperature, and therefore it cannot be
neglected.  As Si has both inversion symmetry and time-reversal
symmetry, all bands are doubly-degenerate across the BZ.  When moving
away from $\Gamma$ there are two (doubly-degenerate) bands with
different effective masses, called ``light-hole'' and ``heavy-hole''
bands.

The self-consistent potential of Si was calculated in the
local-density approximation (LDA) with the Perdew-Wang
parametrization~\cite{PerdewWang1992}, using the {\sc cpw2000}
pseudopotential plane-wave code~\cite{cpw2000}. The pseudopotential
was relativistic Troullier-Martins~\cite{TroullierMartins1991-I,atom}
with a core radius of 1.8 atomic units, ground-state configuration,
and $s$, $p$ and $d$ channels.  The local potential was a smoothed
maximum of all channels, and the nonlocal part was converted to the KB
form.  The self-consistent calculation used a kinetic energy cutoff of
20~Ha for the plane-wave expansion, and a $6 \times 6 \times 6$
uniform grid for BZ integration (28 ${\bf k}$ points,
once symmetry was
taken into account).  The lattice constant in the calculations was
$5.4015\,\,\text{\AA}$.  The lattice constant, energy cutoff,
exchange-correlation functional, and BZ sampling were chosen to be the
same as those of Ref.~\onlinecite{Gonze_mass_2016}, so that any
differences in results could be attributed to the difference in
electronic-structure method: PAW in
Ref.~\onlinecite{Gonze_mass_2016}, versus pseudopotential in the
present work.  The calculations were well converged: increasing cutoff
or BZ sampling had typically a minor effect on the 4th decimal place
of the effective masses.  Using other LDA parametrizations, or using a
different core radius, also only had an effect on the 4th decimal
place.  Using a generalized gradient approximation (GGA) only slightly
changed the third decimal place.  However, using the experimental
lattice constant changed the effective masses by a few percent.  As we
will be checking numerical accuracy, in some tables we report a high
number of decimal places.

\begin{figure} 
\centering
\includegraphics[width=0.5\columnwidth]{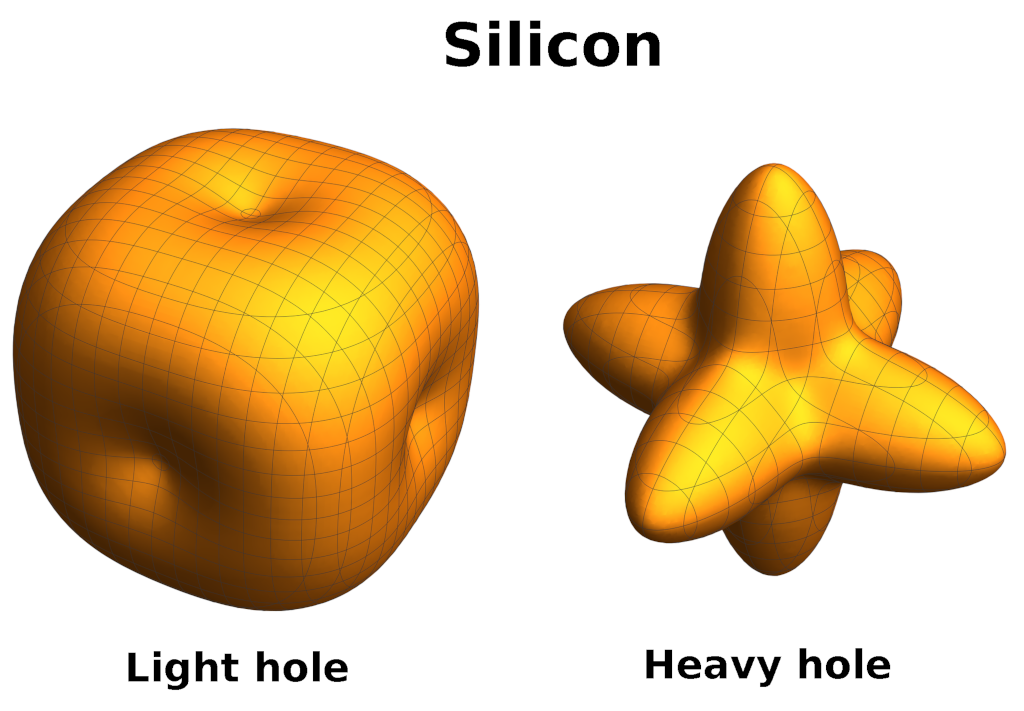}
\caption{The dependence on direction of the second derivative of the
  band energy (inverse effective mass) is shown for the top of the
  valence band of Si at the $\Gamma$ point.  The distance from the center
  of each figure to the surface is proportional to the second
  derivative of the band energy in that direction.  The inverse
  effective masses are clearly anisotropic, and cubic symmetry is
  respected.  The scale of the two figures is not the same.}
\label{fig:Si_polar}
\end{figure}

Away from band degeneracies, effective masses are described by the
Cartesian tensor of second-order energy derivative,
Eq.~\eqref{eq:massnondeg}.  That real-symmetric tensor is described by
its principal axes and inverse masses along those axes.  In a cubic
system and for a nondegenerate state at $\Gamma$, the tensor would
just be a multiple of the identity, with identical effective masses in
every direction.  In Si, the split-off hole band and the lowest
conduction band at $\Gamma$ only have the spin double degeneracy and
have isotropic effective masses.  However, at the top of the valence
band where the bands are four-fold degenerate, the formulation of
section~\ref{subsec:massdirection} must be employed.  The resulting
effective masses are strongly anisotropic, as can be seen in
Fig.~\ref{fig:Si_polar}.  Similar figures can be found in the
literature~\cite{Band_warp_Resca2014}.

Table~\ref{table:Si_mass} shows the effective band mass in Si at the
$\Gamma$ point for high-symmetry directions.  States usually described
as holes appear with negative masses.  States are identified by
symmetry, with the added subscript $v$ for valence or $c$ for
conduction.  The first column contains the results of the present
work.  They agree within a few percent with the PAW calculations shown
in the second column.  Effective masses can also be calculated by
finite differences from energy bands calculated along a few points in
the chosen direction, and we show in that table the results from
Lagrange interpolation of order $n=6$ and $8$, with
$n+1$ points centered at $\Gamma$ and spaced by $\delta = 10^{-3}$,
$10^{-4}$ or $10^{-5}$ (in atomic units).
Instead of an analytical
finite-difference formula for the second derivatives, we use a
recursive algorithm, which has the advantage of being
numerically stable.  As an added bonus, by comparing the second
derivatives at order $n-1$ of the recursion with the final value, an
indication of the reliability of the estimate is
obtained~\cite{poly_interp}.  One can see that for $\delta = 10^{-4}$,
the $n=6$ and $n=8$ interpolations give the same values for the
effective masses as perturbation theory.  For the larger
$\delta = 10^{-3}$ spacing of the sampling wavevectors, there are some
deviations in mass values, but they are still sufficiently
accurate for the physics of the problem.  For the smaller
$\delta = 10^{-5}$ spacing, the numerical instability of Lagrange
interpolation with closely-spaced points
appears, and the results are less reliable.  In fact,
for that small spacing, the degeneracy of the mass values (they should
all be doubly degenerate) is numerically broken, and the values
reported in the table are their mean values.  The main point of the
table is that the implemented perturbative method works and is
very stable and accurate.

\begin{table}
\centering
\resizebox{\textwidth}{!}{%
\begin{tabular}{lcrrrrrrrr}
\hline
\hline
       &     &   \multicolumn{2}{c}{Perturbation} & \multicolumn{5}{c}{Finite differences}  \\
                      &       &    PW-PT    &    PAW-PT  & FD $6,10^{-3}$ & FD $6,10^{-4}$ & FD $8,10^{-4}$  & FD $6,10^{-5}$  & FD $8,10^{-5}$   \\
\hline
$\Gamma_\text{6v}$    &       &  1.161527   &  1.161530  &    1.161655    &    1.161529    &    1.161528     &    1.161640     &    1.161644      \\[.2cm]
$\Gamma_\text{7v}$    &       & -0.226749   & -0.222588  &   -0.226746    &   -0.226749    &   -0.226749     &   -0.226747     &   -0.226747      \\[.2cm]
$\Gamma_\text{8v}$ LH & [100] & -0.191152   & -0.188252  &   -0.191151    &   -0.191152    &   -0.191152     &   -0.191162     &   -0.191162      \\
                      & [110] & -0.139122   & -0.136672  &   -0.139122    &   -0.139122    &   -0.139122     &   -0.139126     &   -0.139280      \\
                      & [111] & -0.132074   & -0.129739  &   -0.132073    &   -0.132074    &   -0.132074     &   -0.132073     &   -0.132225      \\[.2cm]
$\Gamma_\text{8v}$ HH & [100] & -0.260038   & -0.253933  &   -0.260032    &   -0.260038    &   -0.260038     &   -0.260034     &   -0.260034      \\
                      & [110] & -0.529353   & -0.517250  &   -0.529345    &   -0.529353    &   -0.529353     &   -0.529334     &   -0.528980      \\
                      & [111] & -0.664230   & -0.648382  &   -0.664234    &   -0.664230    &   -0.664230     &   -0.664365     &   -0.663356      \\[.2cm]
$\Gamma_\text{6c}$    &       &  0.395668   &  0.385387  &    0.395671    &    0.395669    &    0.395669     &    0.395692     &    0.395694      \\
\hline
\hline
\end{tabular} }
\caption{Effective masses (in atomic units) of Si bands at $\Gamma$.
  The values calculated in the present work with plane waves and
  perturbation theory (PW-PT) are compared with the published values
  obtained using the {\sc abinit} code with projected augmented waves
  and perturbation theory~\cite{Gonze_mass_2016} (PAW-PT), and with
  our finite-difference (FD) estimates.  The finite-difference values
  are from Lagrange interpolation; the $n$ in the notation ``FD n''
  indicates the order of the polynomial and is followed by the value
  (in atomic units)
  of the spacing between the $n+1$ interpolation
  points.}
\label{table:Si_mass}
\end{table}

There is an approximation from the early days of electronic structure
theory, the ${\bf k}\cdot{\bf p}$ method, that uses the second-order
expansion of the Hamiltonian implicit in Eqs.~\eqref{eq:Horder0},
\eqref{eq:Horder1} and~\eqref{eq:Horder2}.  (Obviously, the early work
did not have the projectors of the nonlocal
pseudopotential.)  Given a reference wavevector ${\bf k}_0$ and its
cell-periodic eigenfunctions $\vert {u}_{d n {\bf k}_0} \rangle$, one
can construct an approximate Hamiltonian matrix for a neighboring
wavevector ${\bf k}$,
\begin{equation}
   \begin{aligned}
       H^{{\bf k}\cdot{\bf p}}_{d n,d^\prime n^\prime} & = \langle {u}_{d n {\bf k}_0} \vert H_{{\bf k}_0}  \vert {u}_{d^\prime n^\prime {\bf k}_0} \rangle
                                      + \sum_\alpha ({\bf k} - {\bf k}_0)_\alpha
                                            \langle {u}_{d n {\bf k}_0} \vert H_{{\bf k}_0}^\alpha \vert {u}_{d^\prime n^\prime {\bf k}_0} \rangle   \\
                               & \qquad\qquad + \sum_{\alpha\beta} ({\bf k} - {\bf k}_0)_\alpha
                                        \langle {u}_{d n {\bf k}_0} \vert  H_{{\bf k}_0}^{\alpha\beta}
                                               \vert {u}_{d^\prime n^\prime {\bf k}_0} \rangle  ({\bf k} - {\bf k}_0)_\beta,
   \end{aligned}
   \label{eq:Hkdotp}
\end{equation}
where indices $n$ and $n^\prime$ span a truncated set of bands.  Here
the Luttinger-Kohn basis set~\cite{LuttingerKohn} is used, and the
approximation gets its name from the
${\bf k}\cdot{(-i {\boldsymbol\nabla})}$ factor present in the second
term of Eq.~\eqref{eq:Hk}, from which Eq.~\eqref{eq:Hkdotp} follows.
These are the same ingredients as the perturbation theory described
previously; what is missing is the Sternheimer equation that corrects
for the band truncation.  The advantage of this scheme is that once
the matrix elements in the basis set are calculated for ${\bf k}_0$,
it is trivial to construct $ H^{{\bf k}\cdot{\bf p}}$ for any
${\bf k}$ and, since it is a small matrix, diagonalize it.
Table~\ref{table:Si_k.p} compares the effective masses from
perturbation theory with those obtained from finite differences with a
${\bf k}\cdot{\bf p}$ Hamiltonian based on basis sets with 9, 15, 59,
112 and 259 functions (not counting spin).  One can see that the
15-band model (30 functions counting spin) already describes with
reasonable accuracy (within a few percent) the effective masses.  As
expected, their values converge to the correct ones with increasing
model size.

\begin{table}
\centering
\begin{tabular}{lcrrrrrrr}
\hline
\hline
       &     &   \multicolumn{1}{c}{\textrm{Perturbation}} & \multicolumn{5}{c}{${\bf k}\cdot{\bf p}$}  \\
                      &       &  \multicolumn{1}{c}{\textrm{PW-PT}}
                                         & $n = 9$ & $n = 15$ & $n = 59$ & $n = 112$ & $n = 259$  \\
\hline
$\Gamma_\text{6v}$    &       &  1.161   &  1.082  &  1.081   &  1.125   &  1.141    &  1.158     \\[.2cm]
$\Gamma_\text{7v}$    &       & -0.227   & -0.249  & -0.232   & -0.230   & -0.228    & -0.227     \\[.2cm]
$\Gamma_\text{8v}$ LH & [100] & -0.191   & -0.221  & -0.196   & -0.194   & -0.192    & -0.191     \\
                      & [110] & -0.139   & -0.143  & -0.141   & -0.140   & -0.140    & -0.139     \\
                      & [111] & -0.132   & -0.135  & -0.134   & -0.133   & -0.133    & -0.132     \\[.2cm]
$\Gamma_\text{8v}$ HH & [100] & -0.260   & -0.268  & -0.267   & -0.263   & -0.262    & -0.261     \\
                      & [110] & -0.529   & -0.619  & -0.562   & -0.544   & -0.536    & -0.532     \\
                      & [111] & -0.664   & -0.742  & -0.717   & -0.687   & -0.675    & -0.668     \\[.2cm]
$\Gamma_\text{6c}$    &       &  0.396   &  0.243  &  0.381   &  0.388   &  0.391    &  0.394     \\
\hline
\hline
\end{tabular}
\caption{Effective masses (in atomic units) of Si bands at $\Gamma$.
  The values calculated with plane-wave perturbation theory (PW-PT)
  are compared with those obtained by ${\bf k}\cdot{\bf p}$ models
  using a basis $n = 9$, 15, 59, 112, 259 orbitals (not counting
  spin).}
\label{table:Si_k.p}
\end{table}

\subsection{Gallium arsenide}

\begin{figure}  
\centering
\includegraphics[width=0.8\columnwidth]{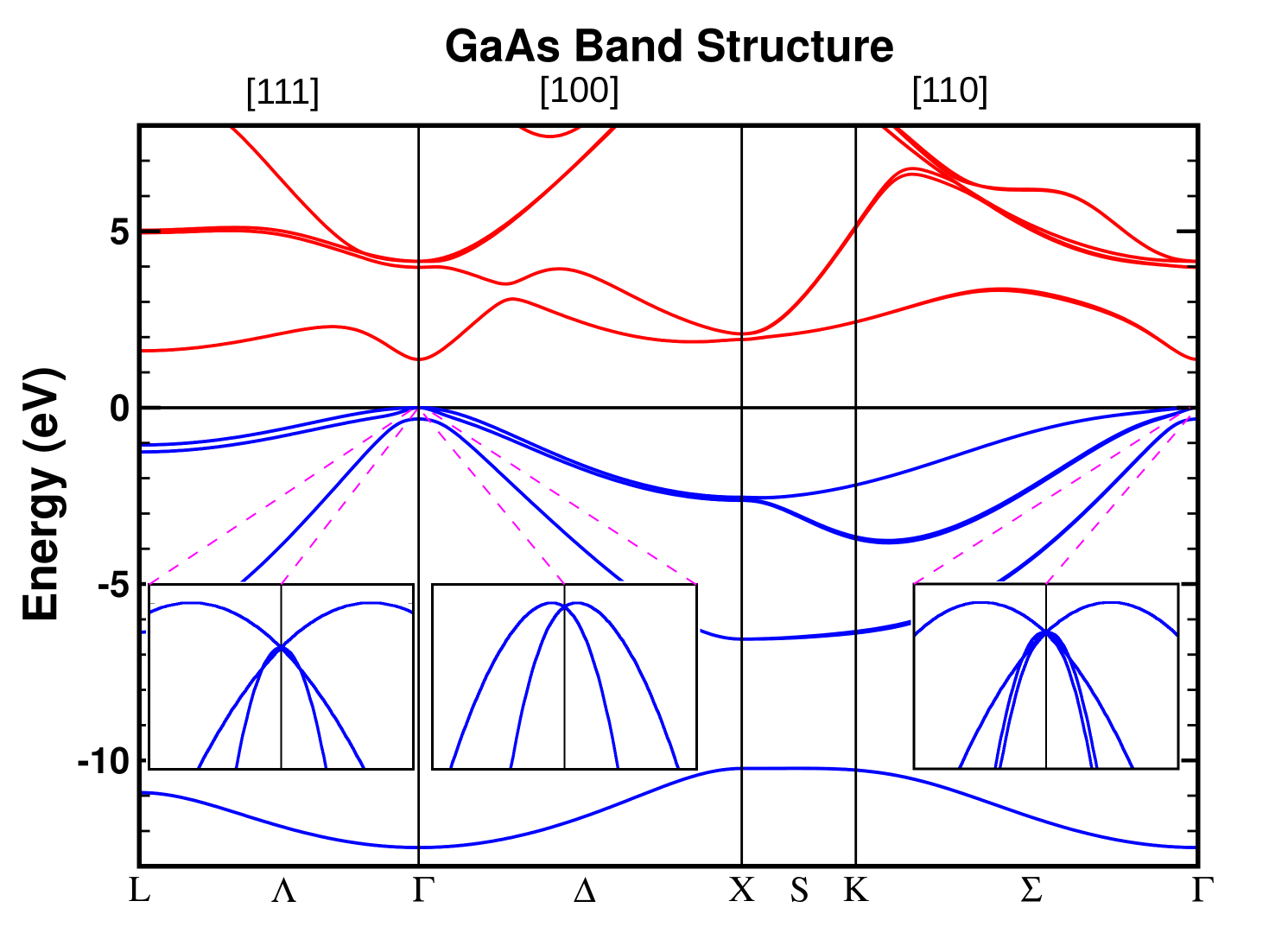}
\caption{The valence bands (blue) and conduction bands (red) of GaAs
  are shown for the main symmetry directions of the fcc lattice.  The
  highly magnified behavior of the hole bands near the top of the
  valence band is shown in the insets, where the energy range is
  $10^{-5}\,\,\text{eV}$ and the wavevector range is
  $1.18 \times 10^{-3}$ atomic units.  The magenta lines connect the
  insets to the regions of the main plot that were magnified.}
\label{fig:GaAsbands}
\end{figure}

While the general features of the gallium arsenide band structure are
similar to those of silicon, the absence of spatial inversion symmetry
means that the bands are not necessarily degenerate, and at the top of
the valence band at $\Gamma$, the band slope may be non-zero.  These
effects are easily missed in a band-structure plot showing the full
energy range of the valence and lower conduction bands, see
Fig~\ref{fig:GaAsbands}, or even on the scale of room
temperature; however, they are clearly seen on the
$10^{-5}\,\,\text{eV}$ energy range of the insets in
Fig~\ref{fig:GaAsbands}.  Perturbation theory at the
$\Gamma$ point will give the first and second derivatives of the
energy at that point, which are
connected with the smallest of the
energy scales.  This behavior near the top of the valence band of GaAs
was described by E.~Dresselhaus in a seminal
paper~\cite{Dresselhaus_ZB}.

The self-consistent potential of GaAs was calculated with the {\sc
  cpw2000} pseudopotential plane-wave code using a modified
Becke-Johnson meta-GGA functional~\cite{TranBlaha2009} that gives an
energy gap close to experiment.  Both Ga and As pseudopotentials were
relativistic, constructed with the Troullier-Martins
recipe~\cite{TroullierMartins1991-I,atom} from a ground-state
configuration and LDA exchange and correlation.  The Ga
pseudopotential had a core radius of 2.5 atomic units for the $s$ and
$p$ channels, and 2.69 atomic units for the $d$ channel.  The As
pseudopotential had a core radius of 2.1 atomic units for the $s$, $p$
and $d$ channels.  In both cases the $d$ channel was used as the local
part, and the nonlocal part was converted to the KB form.  The
self-consistent calculation used a kinetic energy cutoff of 16~Ha for
the plane-wave expansion, and a $4 \times 4 \times 4$ uniform grid for
BZ integration (10 ${\bf k}$ points, once symmetry was taken into
account).  The lattice constant in the calculations was
$5.653\,\,\text{\AA}$.

\begin{table}
\centering
\begin{tabular}{lcrrrrrrr}
\hline
\hline
       &     &   \multicolumn{1}{c}{\textrm{Perturbation}} & \multicolumn{2}{c}{${\bf k}\cdot{\bf p}$} & \multicolumn{3}{c}{Finite differences}  \\
                      &       &    \multicolumn{1}{c}{\textrm{PW-PT}}
                                          & $n = 9$ & $n = 15$ & FD $3,10^{-3}$ & FD $3,10^{-4}$ & FD $3,10^{-5}$ \\
\hline
$\Gamma_\text{6v}$    &       &  1.345679 &  0.837  &  0.838   &  1.345746      &  1.345331      &  1.345676      \\[.2cm]
$\Gamma_\text{7v}$    &       & -0.191713 & -0.222  & -0.195   & -0.191746      & -0.191717      & -0.191714      \\[.2cm]
$\Gamma_\text{8v}$ LH & [100] & -0.152844 & -0.181  & -0.161   & -0.095675      & -0.152846      & -0.152845      \\
                      & [110] & -0.088951 & -0.098  & -0.093   & -0.086067      & -0.099610      & -0.088952      \\
                      & [111] & -0.083548 & -0.091  & -0.088   & -0.083547      & -0.083548      & -0.083548      \\[.2cm]
$\Gamma_\text{8v}$ HH & [100] & -0.152844 & -0.181  & -0.161   & -0.380019      & -0.152846      & -0.152845      \\
                      & [110] & -0.542557 & -1.194  & -0.593   & -0.683390      & -0.328346      & -0.542561      \\
                      & [111] & -0.895971 &  ---    & -0.996   & -0.895868      & -0.895949      & -0.896110      \\[.2cm]
$\Gamma_\text{6c}$    &       &  0.075658 &  0.074  &  0.074   &  0.075646      &  0.075659      &  0.075658      \\
\hline
\hline
\end{tabular}
\caption{Effective masses (in atomic units) of the GaAs bands at
  $\Gamma$. The values calculated with plane-wave perturbation theory
  (PW-PT) are compared with those obtained by ${\bf k}\cdot{\bf p}$
  with $n = 9$ or 15 orbitals (not counting spin), and by finite
  differences (FD). The notation ``FD n'' is the same as in
  Table~\ref{table:Si_mass}.}
\label{table:GaAs_mass}
\end{table}

The calculated effective masses at $\Gamma$ are shown in
Table~\ref{table:GaAs_mass}.  There is an excellent agreement (about
six decimal places) between the masses calculated from perturbation
theory and those obtained from an interpolation of band energies (with
a sixth-order polynomial, and points equally spaced by $10^{-5}$ and
centered in $\Gamma$).  However, in the interpolation, the bands have
to be carefully ``disentangled'' to follow the correct branch after
each crossing.  Even with the help of the quality estimate of the fit
that flags most mistakes~\cite{poly_interp}, that disentanglement is a
very tedious and error-prone job.

\begin{figure}  
\centering
\includegraphics[width=0.9\columnwidth]{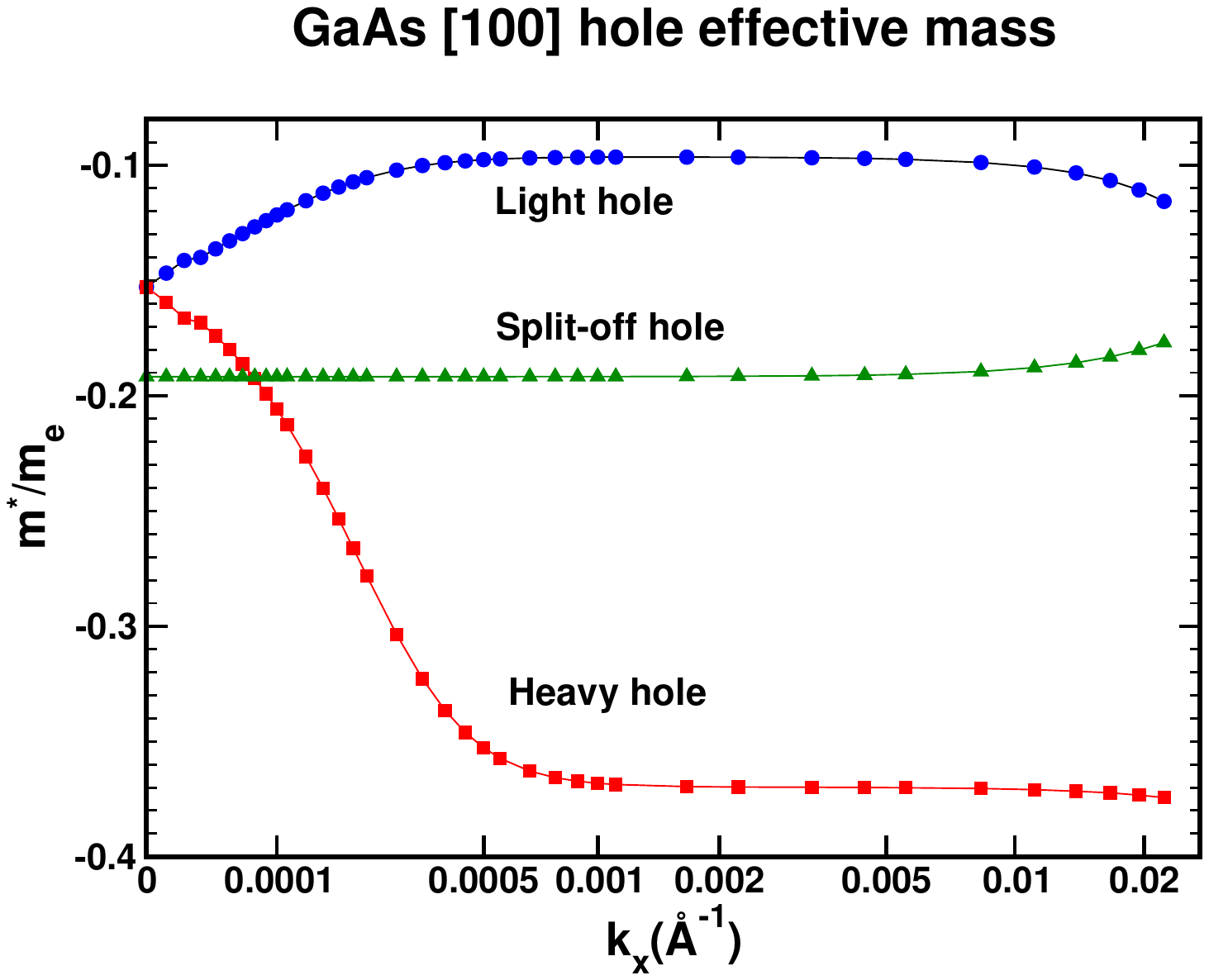}
\caption{The effective masses near the top of the valence bands of
  GaAs are shown as a function of the deviation from the $\Gamma$
  point.  The horizontal scale is ``shifted logarithmic'' as described
  in the text, since the system has several energy and wavevector
  scales.  The split-off hole mass is almost constant.  The light- and
  heavy-hole masses are constant over a sizeable wavevector range, but
  converge to the same value at $\Gamma$.}
\label{fig:GaAsmasses}
\end{figure}

A clear sign of the peculiarities in the effective masses is that at
$\Gamma$ and in the $[100]$ direction they are identical for the
``heavy'' and ``light'' holes.  In that direction, the four bands that
are degenerate at $\Gamma$ split into two doubly-degenerate bands that
are mirror images of each other, and therefore have the same second
derivatives at $\Gamma$.  From the insets of Fig.~\ref{fig:GaAsbands},
it is clear that near $\Gamma$ the bands deviate from parabolas.  In
Fig.~\ref{fig:GaAsmasses}, the longitudinal masses along $[100]$ are
plotted as a function of distance from the $\Gamma$ point.  The
reciprocal vector is on a scale $x = \log(1+k/k_0)$ that is linear
near the origin ($k \ll k_0$) but logarithmic far from it
($k \gg k_0$).  The value used in the plot is
$k_0 = 10^{-4}\,\,\text{\AA}^{-1}$.  The mass of the split-off hole is
almost constant throughout the plot range.  The heavy-hole and
light-hole masses are identical at $\Gamma$, deviating linearly at
very small $k$ until they settle at an almost constant value.

The GaAs bands near $\Gamma$ illustrate how the degenerate
perturbation theory described in the previous section works in
practice.  In the $[111]$ direction ($\Gamma-$L), diagonalization of
the $4 \times 4$ matrix
$A^{(1)}(\hat{\bf q}=[111],n=\text{VBM},{\bf k}=\Gamma)$ of
Eq.~\eqref{eq:firstorderE} for the four-fold degenerate valence band
maximum (VBM) gives two nonzero and nondegenerate band velocities with
opposite signs, and a twice-degenerate vanishing band velocity (see
left inset of Fig.~\ref{fig:GaAsbands}).  For the two nondegenerate
band velocities, the masses are obtained from the relevant diagonal
elements of $\tilde A^{(2)}([111],\text{VBM},\Gamma)$, which results
from applying to $A^{(2)}([111],\text{VBM},\Gamma)$ the unitary
transformation that diagonalizes $A^{(1)}([111],\text{VBM},\Gamma)$;
see Eqs.~\eqref{eq:A2dd} to \eqref{eq:secondderiv}.  These two
diagonal elements are identical, implying that the corresponding
masses are the same.
For the doubly-degenerate
level with vanishing band velocity, the masses are obtained by
diagonalizing the corresponding $2 \times 2$ submatrix of
$\tilde A^{(2)}([111],\text{VBM},\Gamma)$, yielding two identical
masses.  In the $[100]$ direction ($\Gamma-$X),
$A^{(1)}([100],\text{VBM},\Gamma)$ has two doubly-degenerate nonzero
band velocities with opposite signs (see middle inset of
Fig.~\ref{fig:GaAsbands}).  For each sign, the corresponding
$2 \times 2$ submatrix of $\tilde A^{(2)}([100],\text{VBM},\Gamma)$
must be diagonalized; the final result is four identical masses as
discussed previously in connection with Fig~\ref{fig:GaAsmasses}.
Finally, in the $[110]$ direction ($\Gamma-$K),
$A^{(1)}([110],\text{VBM},\Gamma)$ has four distinct band velocities:
two with large absolute values, and two with small absolute values
(see right inset of Fig.~\ref{fig:GaAsbands}).  All the masses are
obtained from the corresponding diagonal elements of
$\tilde A^{(2)}([110],\text{VBM},\Gamma)$. The fact that at $\Gamma$
the effective mass degeneracies are multiples of two is a consequence
of time-reversal symmetry.

\subsection{Comparison between perturbation theory and finite differences}

Since effective masses can be calculated by finite differences or by
perturbation theory, the advantages and disadvantages of each approach
should be compared.  Finite differences can be used with any code that
is able to produce a band structure.  Perturbation theory, on the
other hand, needs an implementation like the one described here and in
previous works~\cite{Gonze_mass_2016,Blaha_mass_2021}.  However, as
can be seen from Tables~\ref{table:Si_mass} and~\ref{table:GaAs_mass},
the best parameters for the finite differences are not obvious.  If
one inspects Fig.~\ref{fig:GaAsmasses}, it is clear that using a
spacing between sampling points of $10^{-5}\,\,\text{\AA}$ will yield
a very different result from a spacing of $10^{-3}\,\,\text{\AA}$.
Furthermore, when one has a situation like the one depicted in the
insets of Fig.~\ref{fig:GaAsbands}, and only a few sampling points are
used, it is not trivial how to ``connect the dots'' in the correct
way.  Perturbation theory does not need any such guesswork.

The computationally expensive step in perturbation theory is
solving the Sternheimer equation.  When an iterative algorithm is
used, the [time-consuming] operation is applying
the Hamiltonian to a trial solution.  This is the same operation as in
an iterative diagonalization of the Hamiltonian to obtain the
eigenstates.  As the number of iterative steps is of the same order of
magnitude, perturbation theory has a limited overhead and, more
importantly, it scales with system size in the same way as the
iterative diagonalization.  Finite differences has the same scaling
with system size, but as it needs several Hamiltonian diagonalizations
for the {\bf k}-point sampling, it is computationally more expensive.

\section{Gapped graphene}
\label{sec:graphene}

Gapped graphene is a system where the inversion symmetry of the
two-dimensional pristine structure is broken by an external potential,
typically associated with a substrate on which graphene is grown
epitaxially~\cite{graphene_epitaxial_2007,review_graphene_Neto_Peres_2009}.
The equivalence between the two atomic sites in the primitive unit
cell is broken, and as a result a gap opens in the Dirac cones at the
Fermi level.  The simplest tight-binding (TB)
model~\cite{graphite_TB_Wallace_1947,graphene_SO_Kane_Mele_2005}, with
just one $p_z$ orbital per site, leads to a two-band low-energy model
with just two parameters: the magnitude $\Delta$ of the gap
 and the Fermi velocity
$v_\text{F} = \sqrt{3} a t / 2$~\cite{graphite_TB_Wallace_1947},
with $a$ the lattice constant and $t$
the hopping integral of the TB model.  Quantum-geometric quantities
have simple analytical expressions in two-band
models~\cite{pozo-prb20,graf-prb21}, and the low-energy model for
gapped graphene has been widely used in recent times to investigate
such quantities~\cite{XiaoYaoNiu2007}.

Adding a bespoke external potential with a gaussian shape to one of
the atomic sites in a first-principles calculation for graphene opens
such a gap in the Dirac cones.  This allows a comparison of the
quantum-geometric quantities calculated from first principles with
those from the low-energy model.  Obviously, one cannot expect the
exact same results, but they should be similar, particularly in the
limit of a vanishing gap.

The self-consistent potential of broken-symmetry graphene was
calculated using a modified Becke-Johnson meta-GGA
functional~\cite{TranBlaha2009} in a supercell configuration.  The
Tran-Blaha constant~\cite{TranBlaha2009} was fixed at
$c_\text{TB} = 1.04$, since the original recipe is inadequate for
isolated 2D materials.  While the LDA approximation is known to
underestimate the bonding energy of $\sigma$ electrons with respect to
$\pi$ electrons~\cite{Schabel_graphite}, we find that the modified
Becke-Johnson meta-GGA functional gives results closer to experiment.
A Troullier-Martins~\cite{TroullierMartins1991-I,atom} pseudopotential
was constructed from a local-density ground-state configuration, with
a core radius of 1.3 atomic units for the $s$, $p$
and $d$ channels.  The local potential was a smoothed maximum of all
channels, and the nonlocal part was converted to the Kleynman-Bylander
form.  The self-consistent calculation used a kinetic energy cutoff of
50~Ha for the plane-wave expansion, and a $6 \times 6 \times 2$
uniform grid for BZ integration.  As it is a slab calculation, a
$6 \times 6 \times 1$ grid could be used instead, but the number of
irreducible $k$ points would be the same, and the convergence with
slab separation distance would be slightly slower.  The lattice
constant in the calculations was $2.456\,\,\text{\AA}$ on
the graphene plane, three times that value was used in the
perpendicular supercell direction.  The added inversion-breaking
potential was tweaked to open a gap of 0.28~eV; this is the same value
that was used in Ref.~\onlinecite{XiaoYaoNiu2007}, and is also very
close to the experimental value reported for epitaxial graphene on a
SiC substrate~\cite{graphene_epitaxial_2007}.

\begin{figure}  
\centering
\includegraphics[width=0.6\columnwidth]{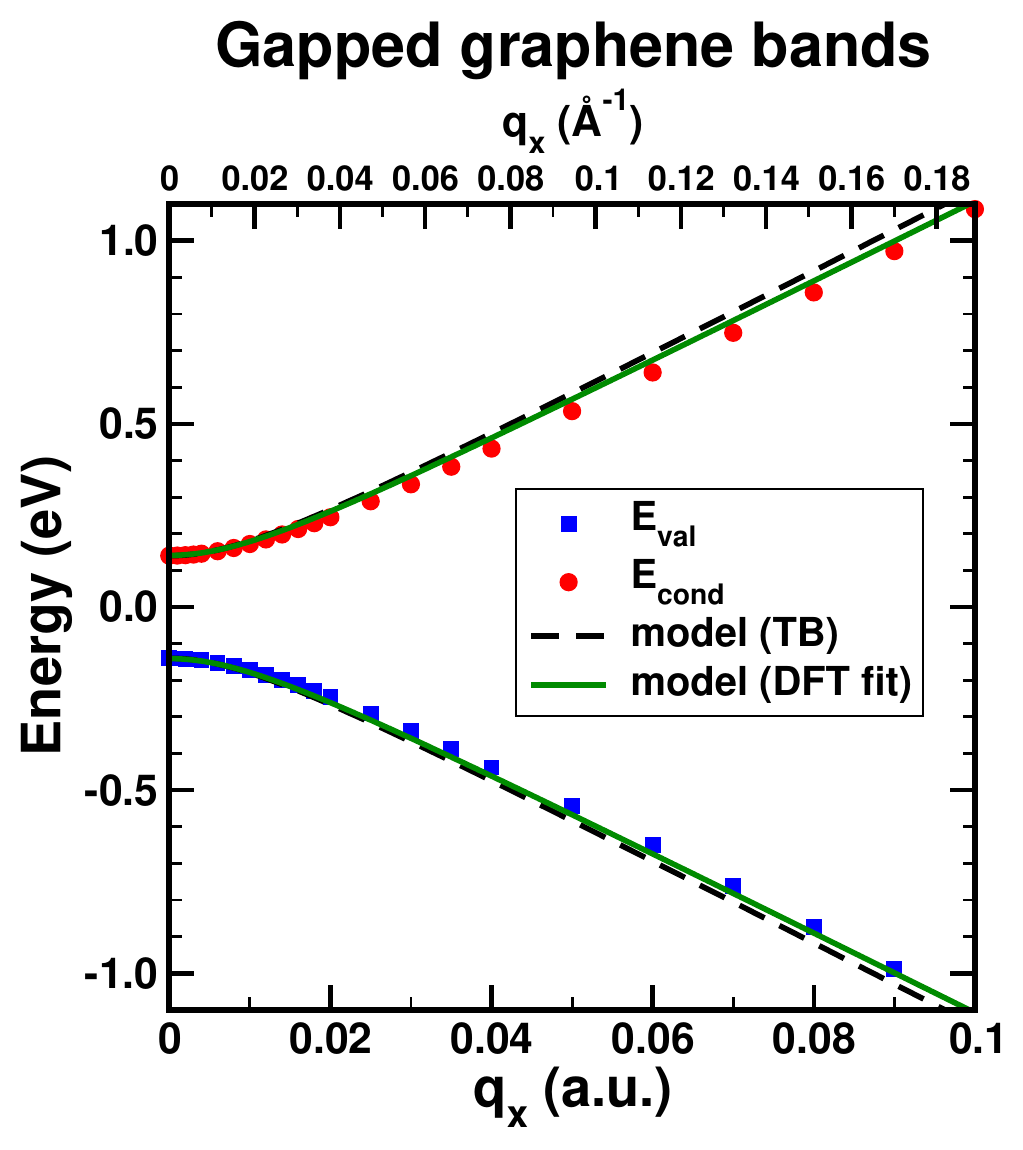}
\caption{The energy bands of gapped graphene are shown as a function
  of wavevector distance from (BZ corner) in the direction of the BZ
  center.  The blue squares and red dots are the valence and
  conduction bands obtained by adding a bespoke potential to the DFT
  calculation.  The dashed-black lines and solid-green lines both
  pertain to the low-energy model, and are obtained by inserting
  slightly different sets of parameters into Eq.~\eqref{eq:twoband}.}
\label{fig:band_energy_graph}
\end{figure}

Figure~\ref{fig:band_energy_graph} shows the DFT-pseudopotential
energy bands as a function of wavevector distance $q_x$ from K towards
$\Gamma$, together with the bands of the low-energy model,
\begin{equation}
E_\pm(q_x,0) =  \pm \frac{\Delta}{2} \sqrt{1+(q_x/q_0)^2},
\label{eq:twoband}
\end{equation}
where we defined a characteristic wavevector,
\begin{equation}
    q_0 = \frac{\Delta}{\sqrt{3} a t} = \frac{\Delta}{2 v_\text{F}}.
\end{equation}
In the paper of Xiao {\it et.\ al}~\cite{XiaoYaoNiu2007} the
parameters are $\Delta = 0.28\,\,\text{eV}$, $a =2.456\,\,\text{\AA}$
and $t = 2.82\,\,\text{eV}$, so that
$q_0 \simeq 0.01235\,\,\text{a.u.}$ With this value of $q_0$, the
energy bands of the low-energy model (dashed-black lines in
Fig.~\ref{fig:band_energy_graph}) look fine for $q_x \lesssim q_0$,
which is not surprising as the external potential was chosen to give
the same gap value $\Delta$ at K.  However, deviations for
$q_x \gg q_0$ are noticeable, meaning that the Fermi velocity of the
underlying pseudopotential graphene energy bands (without the added
potential) is slightly lower than assumed in
Ref.~\onlinecite{XiaoYaoNiu2007}.  If we use the effective mass at K
from the gapped-graphene DFT calculation as the second constraint
in the model (more precisely, the average of the absolute
values of the valence and conduction effective masses), we obtain a
value of $q_0 = 0.01283\,\,\text{a.u.}$ for the characteristic
wavevector, resulting in slightly different bands (solid-green lines
in Fig.~\ref{fig:band_energy_graph}) which are closer to the DFT bands
for $q_x \gg q_0$.  The agreement is not perfect, and should not be
expected to be, but it is very good.

\begin{figure} 
\centering
\includegraphics[width=0.6\columnwidth]{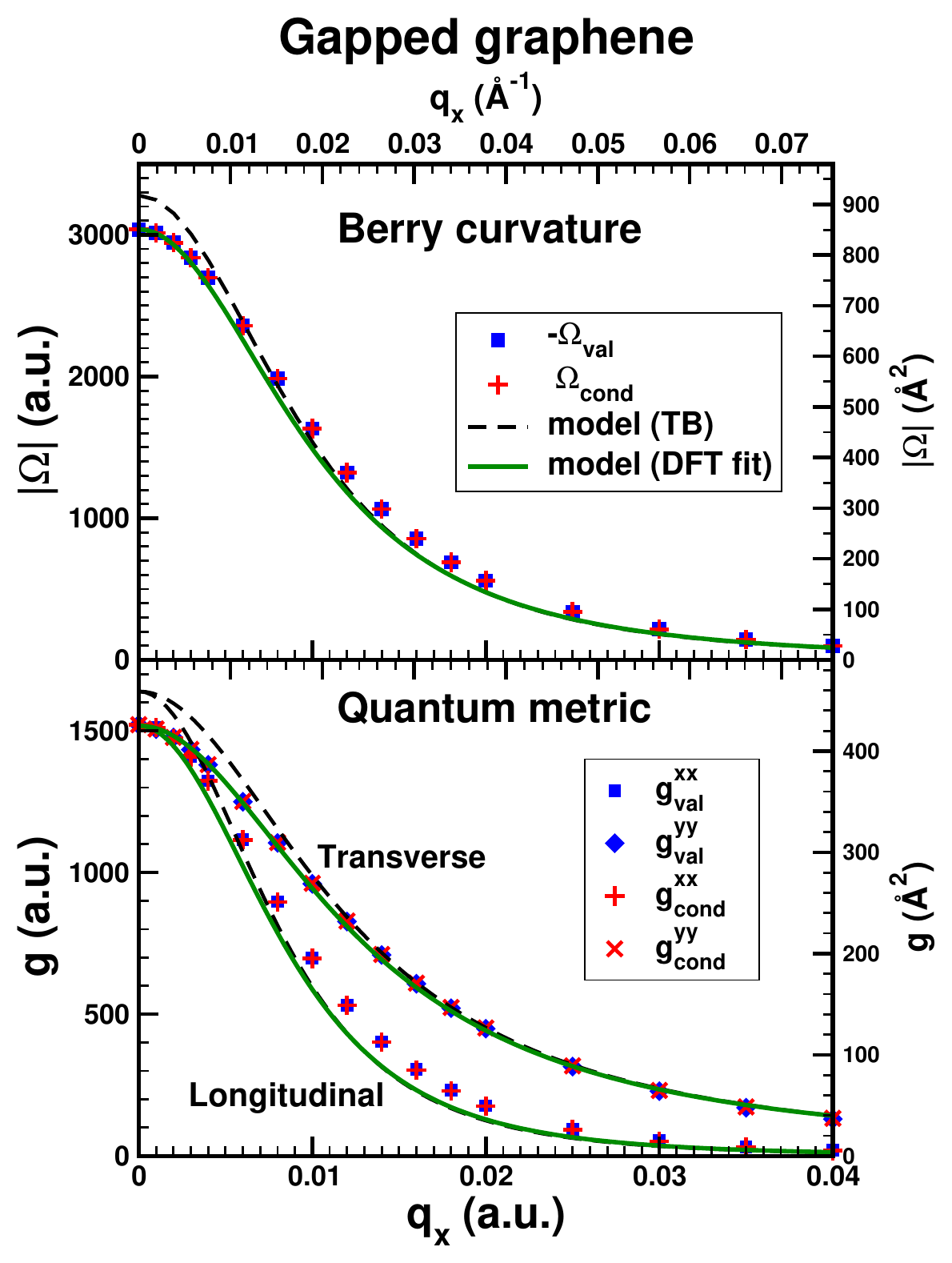}
\caption{The absolute value of the Berry curvature (top panel) and
  components of the quantum metric tensor (bottom panel) of gapped
  graphene are shown as a function of the distance from K towards
  $\Gamma$.  The DFT results are shown using blue symbols for the
  valence band and red symbols for the conduction band.  The
  dashed-black lines and solid-green lines both pertain to the
  low-energy model and are obtained by inserting slightly different
  sets of parameters into Eqs.~\eqref{eq:omega} and~\eqref{eq:qm_g}.}
\label{fig:berry_curvature_metric}
\end{figure}

In the top panel of Fig.~\ref{fig:berry_curvature_metric}, the Berry
curvature $\Omega=\Omega_{xy}$ calculated from DFT is compared with
the low-energy model~\cite{XiaoYaoNiu2007},
\begin{equation}
  \Omega(q_x,0) =  \pm \frac{1}{2 q_0^2} \frac{1}{\left[1+(q_x/q_0)^2\right]^{3/2}};
  \label{eq:omega}
\end{equation}
both quantities are plotted as a function of wavevector distance from
K towards $\Gamma$.  The Berry curvature has opposite signs for the
two bands, and we plot the absolute value to emphasize that on the
scale of the figure, the differences between the two bands are not
visible.  The sign of $\Omega$ alternates between the two K points
that are not equivalent by translation.  It also flips depending on
top of which carbon atom in the unit cell the potential is added to,
or if that potential is attractive or repulsive.  In the low-energy
model, $\Delta$ is often considered as a signed quantity, adding a
further sign choice.  As there is a relationship between the Berry
curvature and the effective masses at $q=0$ in the model [see
Eq.~\eqref{eq:twoband} and Eq.~\eqref{eq:omega}], it is not surprising
that fitting the effective mass at K gives an excellent agreement with
the Berry curvature near that point (solid-green line).  Overall, the
low-energy model can describe quite well the DFT-pseudopotential
results in the displayed range.

The DFT-pseudopotential quantum metric is compared in the bottom panel
of Fig.~\ref{fig:berry_curvature_metric} with that of the low-energy
model, given by Eq.~\eqref{eq:qm_g} below.  By symmetry, the
off-diagonal components are null for our choice of axis orientation
and plot direction, therefore we only plot the two in-plane diagonal
components.  In the low-energy model they are given by
\begin{equation}
  \begin{aligned}
      g^{xx}(q_x,0) & = \frac{1}{4 q_0^2} \frac{1}{\left[1+(q_x/q_0)^2\right]^2}  \\
      g^{yy}(q_x,0) & = \frac{1}{4 q_0^2} \frac{1}{\left[1+(q_x/q_0)^2\right]^1}, \\
  \end{aligned}
  \label{eq:qm_g}
\end{equation}
with $x$ in the direction from K to $\Gamma$.  At K we have
$g^{xx}=g^{yy}=|\Omega|/2$ [note the factor-of-two difference in
Eqs.~\eqref{eq:curvature} and~\eqref{eq:metric}], and the results from
the model with the fitted $q_0$ are very close to the DFT values,
particularly for $q_x \ll q_0$.  As in the case of the Berry
curvature, the differences between the DFT values for the
valence and conduction bands are so small that they are not noticeable
on the scale of the figure.  The dependence on $q_x$ is very different
for the longitudinal and transverse components of the quantum metric.
The model fit of the DFT results is very good for the longitudinal
component, but for the transverse component there is a noticeable
deviation.  As the system has a third dimension, the $zz$ component of
the metric has a nonzero value, $g^{zz} = 0.56\,\,\text{\AA}^2$ at
$q=0$, which is very small on the scale of
Fig.~\ref{fig:berry_curvature_metric}, and depends very weakly on
$q_x$.

\begin{figure} 
\centering
\includegraphics[width=0.6\columnwidth]{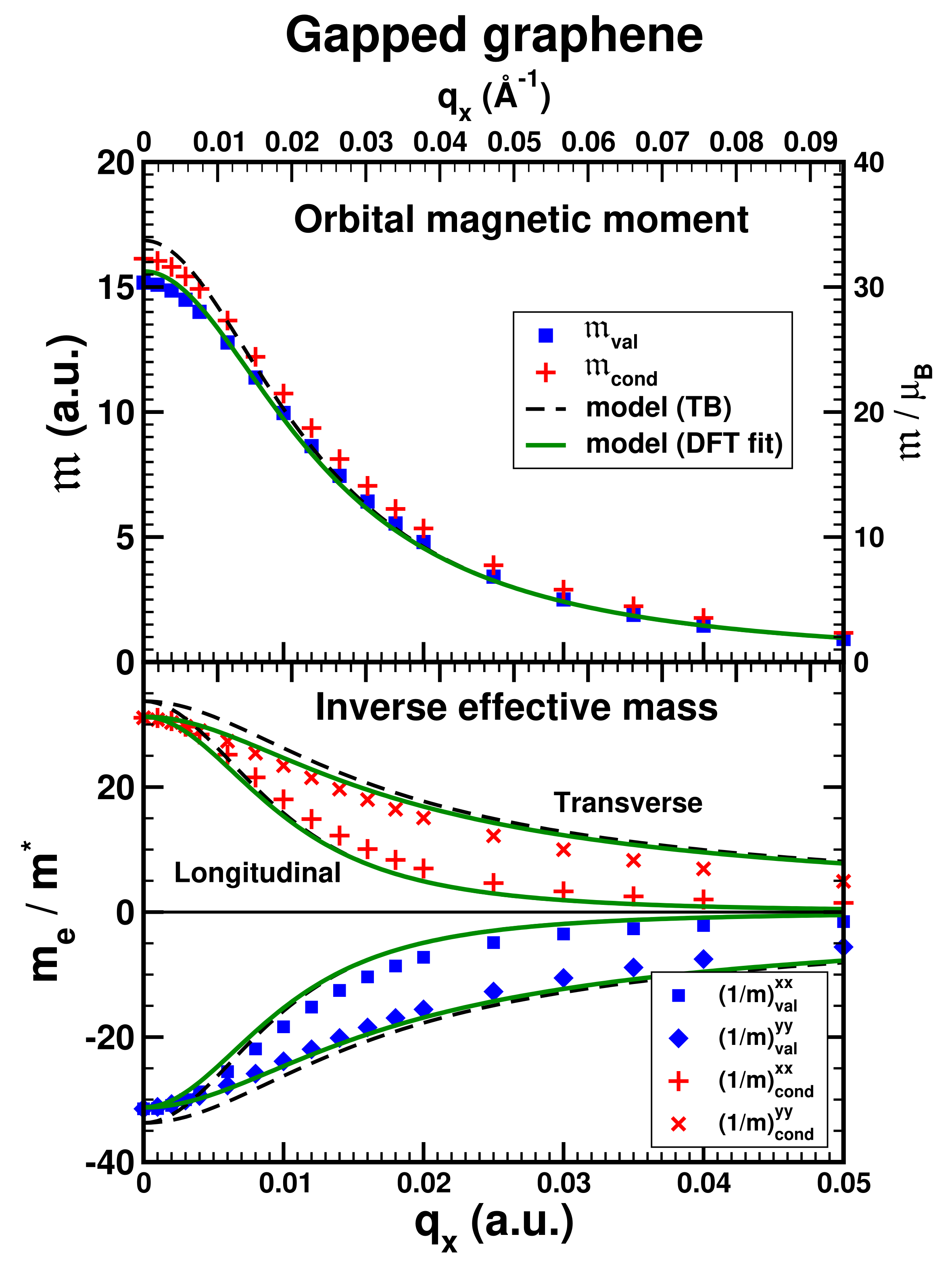}
\caption{The orbital magnetic moments (top panel) and inverse
  effective masses (bottom panel) of gapped graphene are shown as a
  function of the distance from K towards $\Gamma$.  The longitudinal
  components of the inverse effective mass tensor are the second
  derivatives of the energy dispersions in
  Fig.~\ref{fig:band_energy_graph}.  The DFT results are shown using
  blue symbols for the valence band and red symbols for the conduction
  band.  The dashed-black lines and solid-green lines both pertain to
  the low-energy model and are obtained by inserting slightly
  different sets of parameters into Eqs.~\eqref{eq:mass_gr}
  and~\eqref{eq:orb_mag}.}
\label{fig:orbital_mass_graph}
\end{figure}

Turning now to the inverse effective masses, in the bottom panel of
Fig.~\ref{fig:orbital_mass_graph} the DFT results are compared with
those of the low-energy model, which are evaluated as
\begin{equation}
  \begin{aligned}
     \Bigl(\frac{m_\text{e}}{m^*}\Bigr)^{xx}(q_x,0) & = \pm\frac{\Delta}{2 q_0^2}
                  \frac{1}{\left[ 1+(q_x/q_0)^2 \right]^{1/2}}  \\
     \Bigl(\frac{m_\text{e}}{m^*}\Bigr)^{yy}(q_x,0) & = \pm\frac{\Delta}{2 q_0^2}
                  \frac{1}{\left[ 1+(q_x/q_0)^2 \right]^{3/2}},
  \label{eq:mass_gr}
  \end{aligned}
\end{equation}
with $m_\text{e}$ the electron mass (one in atomic units).  We have
omitted from the right-hand side a term $1/a_0^2 E_\text{H}$, with
$a_0$ the Bohr radius and $E_\text{H}$ the Hartree energy, as they are
equal to one in atomic units.  The small deviations seen in
Fig.~\ref{fig:band_energy_graph} between the DFT and model energy
bands get magnified in the inverse masses, which are their second
derivatives.  For $q_x=0$, there is a small difference in masses
between the valence and conduction bands that is not apparent on
the scale of the figure.  As the average effective mass was used for
the value of $q_0$ in the DFT fit, the agreement between the model and
DFT calculations is not surprising.  When moving away from K towards
$\Gamma$, the longitudinal and transverse inverse masses become
different, the longitudinal mass decreasing at a much faster rate.

The final quantum-geometric quantity on our list is the orbital
magnetic moment, which is plotted in the top panel of
Fig.~\ref{fig:orbital_mass_graph}. The results for the low-energy
model were obtained from the following expression,
\begin{equation}
    \mathfrak{m}(q_x,0) = \pm 2\mu_\text{B} \frac{\Delta}{4 q_0^2}
          \frac{1}{\left[ 1+(q_x/q_0)^2 \right ]^1}
    \label{eq:orb_mag}
\end{equation}
where the Bohr magneton, $\mu_\text{B}$, has the value $1/2$ in atomic
units, and we have omitted again from the right-hand side the
$1/a_0^2 E_\text{H}$ factor.  In contrast with previous figures, there
is a clear difference between the orbital magnetic moments of the
valence and conduction bands at K, although they both have values (in
atomic units) close to half the inverse mass, as predicted by the
low-energy model.  The sign of the orbital magnetic moment is the same
for the two bands, but, like the Berry curvature, it depends on which
of the two inequivalent K points is considered, or on which atom the
symmetry-breaking potential is placed.  A consistency check is that
the orbital moment must have the same sign as the Berry curvature of
the conduction band.

\begin{figure} 
\centering
\includegraphics[width=0.6\columnwidth]{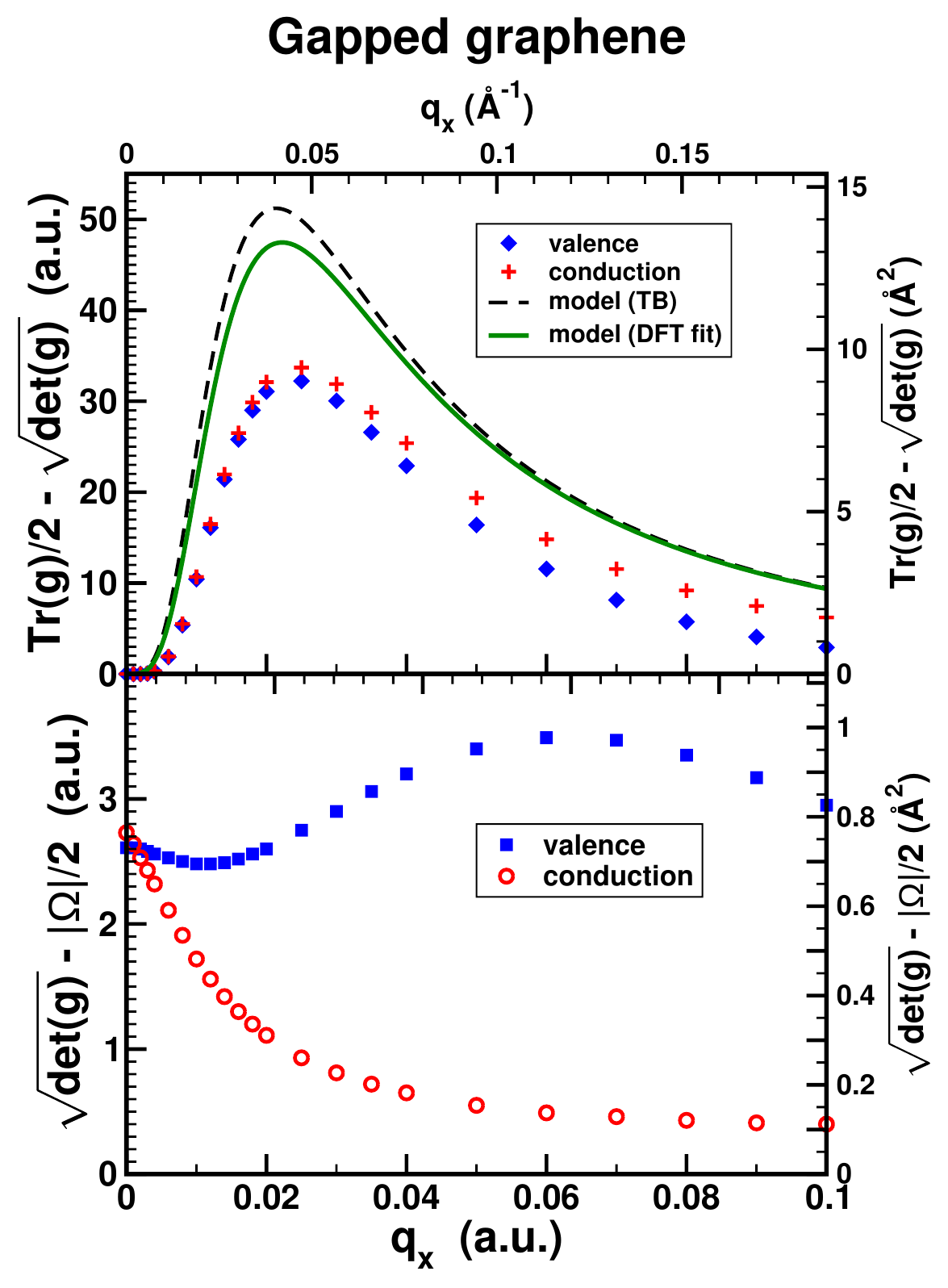}
\caption{The relations in Eq.~\eqref{eq:roy-inequalities}
  between the trace and determinant of the quantum metric $g$ and the
  magnitude of the Berry curvature $\Omega$ are illustrated by
  plotting differences between those quantities as a function of the
  distance from K towards $\Gamma$.  Notice the change of scale with
  respect to Fig.~\ref{fig:berry_curvature_metric}.  In the top panel,
  the dashed-black lines and solid-green lines both pertain to the
  low-energy model, and are obtained by inserting slightly different
  sets of parameters into Eq.~\eqref{eq:qm_g}.  For the quantity
  plotted in the bottom panel, that model predicts a null value.}
\label{fig:ineq_graph}
\end{figure}

Next, we analyze the close connection between the real and imaginary
parts of the quantum-geometric tensor, i.e., between the quantum
metric and the Berry curvature.  For nondegenerate states and
two-dimensional materials, the quantum-geometric tensor becomes a
$2 \times 2$ hermitian matrix.  The three invariants of such a matrix
are the determinant and trace of the real-symmetric part (quantum
metric), and the magnitude of the
imaginary-antisymmetric part (Berry curvature).  From the positive
semi-definiteness of the quantum-geometric tensor, one obtains the
following inequalities between the three
invariants~\cite{roy-prb14,Mera2021},
\begin{equation}
\frac{1}{2}\operatorname{tr} g \geq \sqrt{\det{g}} \geq \frac{1}{2} |\Omega|.
\label{eq:roy-inequalities}
\end{equation}
(The first inequality is just between the arithmetic and geometric
averages of the non-negative eigenvalues of the quantum metric.)
Inspection of Eqs.~\eqref{eq:omega} and~\eqref{eq:qm_g} shows that the
second inequality saturates for the low-energy model (this actually
happens for any two-band model~\cite{Mera2021}), while the first only
saturates in the limits $q_x \ll q_0$ and $q_x \gg q_0$.  In
Fig.~\ref{fig:ineq_graph}, the quantities
$\frac{1}{2}\operatorname{tr} g - \sqrt{\det{g}}$ and
$\sqrt{\det{g}} - \frac{1}{2} |\Omega|$ are plotted as a function of
wavevector distance from K towards $\Gamma$.  Comparing the vertical
scale of this figure with that of
Fig.~\ref{fig:berry_curvature_metric} shows that the inequalities are
very close to being saturated, particularly in the case of
$\sqrt{\det{g}} - \frac{1}{2} |\Omega|$,
where the two-band model would predict a null value.

The two-band low-energy model fails to take into account the presence
of other bands in the DFT calculation, which are responsible for the
small nonzero values displayed in the bottom panel of
Fig.~\ref{fig:ineq_graph}.  The presence of those other bands is also
responsible for the small deviations between DFT and model results
seen in Figs.~\ref{fig:band_energy_graph}
to~\ref{fig:orbital_mass_graph}, but there the difference is only
quantitative and not qualitative.  To estimate the corrections to the
two-band model, note that the sum-over-states expression for
$\QQ_{n {\bf k}} \vert {u}_{d n {\bf k}}^\alpha \rangle$ in
Eq.~\eqref{eq:sternsol} contains an energy-difference denominator.  The
Berry curvature and quantum metric will therefore depend on the
inverse of the square of those energy differences.  The smallest among
them is the gap $\Delta = 0.28\,\,\text{eV}$; the next relevant level
(taking mirror symmetry into account) is about 9~eV away, and
therefore $9 / 0.28 \sim 32$ times farther away.  The corrections to
the metric and Berry curvature should therefore be of the order of
$(1/32)^2$, or about 0.1\%.  Indeed, this is the factor between the
vertical scale of the bottom of Fig.~\ref{fig:ineq_graph} and that of
Fig.~\ref{fig:berry_curvature_metric}.  In the case of the inverse
mass and orbital moment, the sum-over-states expressions depend on the
inverse of the energy differences (not squared); as such, deviations
from the two-band model are expected to be more pronounced, of the
order of 3\%.  That is in fact almost the value by which the orbital
magnetic moments at K in Fig.~\ref{fig:orbital_mass_graph} deviate
from their average.  The difference between the absolute values of the
conduction and valence effective masses is 1.2\%.

\section{Trigonal tellurium}
\label{sec:tellurium}

Tellurium is one of the simplest crystals with interesting
quantum-geometric band properties, which depend on a strong spin-orbit
interaction in a reasonably heavy element, and on its chiral
structure.  As trigonal tellurium has six bands within less than 1~eV
from the gap, it presents a more complex situation than the two bands
of gapped graphene.

The self-consistent potential for the left-handed structure, with
space group P3$_2$21, was calculated with the {\sc cpw2000} code using
a modified Becke-Johnson meta-GGA functional~\cite{TranBlaha2009}.
The pseudopotential was relativistic
Troullier-Martins~\cite{TroullierMartins1991-I,atom} with a core
radius of 2.6 atomic units, ground state configuration, and $s$, $p$
and $d$ channels.  The local potential was a smoothed maximum of all
channels, and the nonlocal part was converted to the KB
form~\cite{KleinmanBylander1982}. The crystal lattice constants were
$a = 4.44\,\,\text{\AA}$ and $c = 5.91\,\,\text{\AA}$, and the
internal displacement parameter was
$u = 0.269$~\cite{Te_Asendorf1957}.  The self-consistent calculation
used a kinetic energy cutoff of 18~Ha for the plane-wave expansion,
and a $6\times 6 \times 6$ uniform grid for BZ integration
(63 ${\bf k}$ points,
once symmetry was taken into account).  The Tran-Blaha
parameter~\cite{TranBlaha2009} was fixed at $c_\text{TB} = 1.10$. This
value was chosen to give a band gap of $0.312\,\,\text{eV}$, the same
as a previous calculation with the HSE06 hybrid
functional~\cite{Tsirkin_Te_2018}, which is close to the
$0.314\,\,\text{eV}$ of a GW calculation~\cite{Te_GW_2015}, and to the
experimental value of $0.323\,\,\text{eV}$~\cite{Te_gap_eremets1977}.

\begin{figure} 
\centering
\includegraphics[width=0.8\columnwidth]{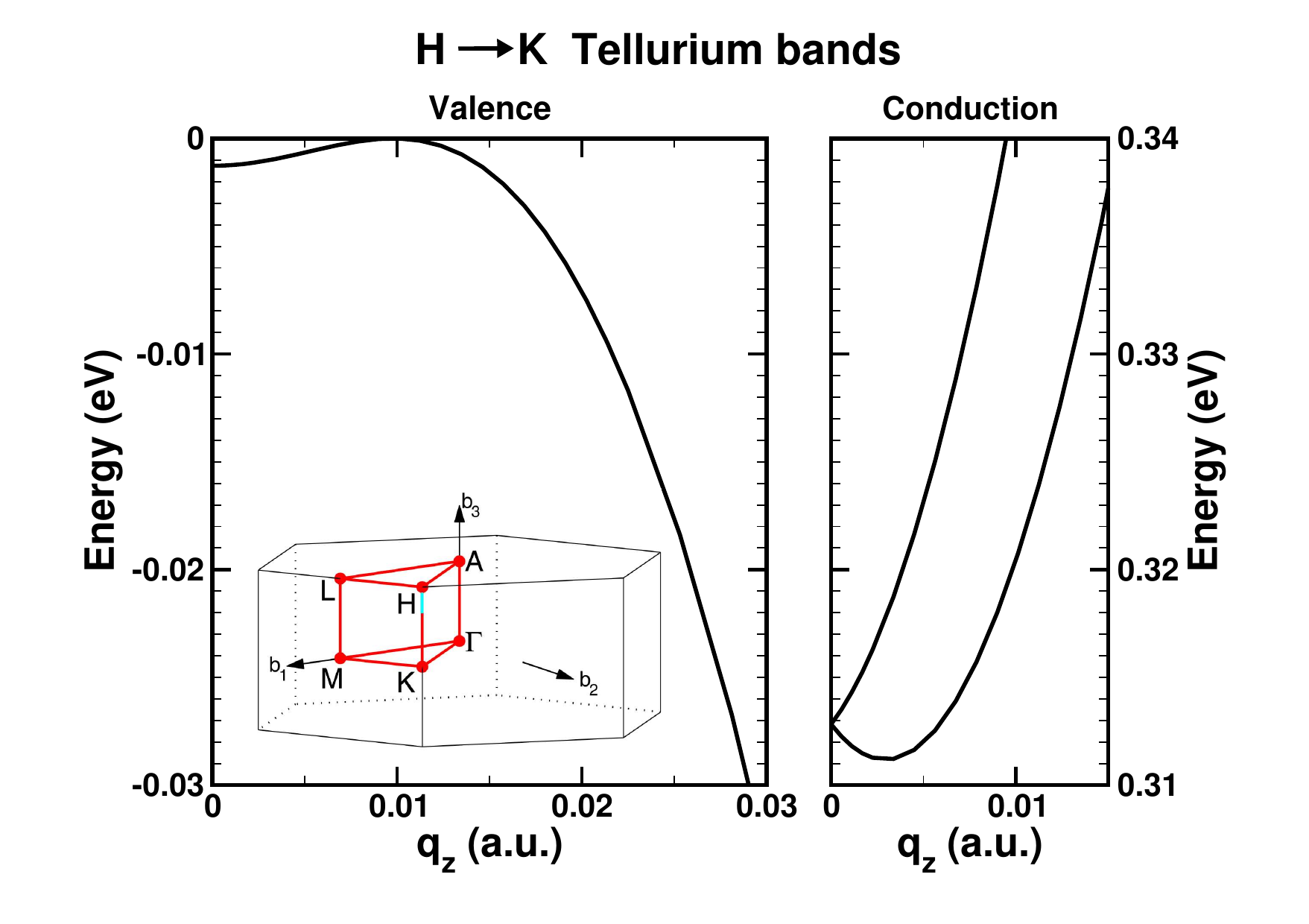}
\caption{The energy bands of Te near the band gap are shown along the
  H--K line, near the H point ($q_z=0$).  The upper valence and lower
  conduction bands are shown in the left and right panels,
  respectively.  On this highly detailed scale, the ``camel-back''
  structure of the valence band is clearly seen.  The inset shows the
  hexagonal Brillouin zone.  Therein, the short cyan segment on the
  H--K line highlights the considered region.}
\label{fig:band_Te}
\end{figure}

The valence-band maximum (VBM) of trigonal Te, relevant for the
naturally $p$-doped material, is located on the H--K line, the lateral
edge of the hexagonal BZ, very close to the H points, which form the
vertices of the hexagonal BZ: see inset of Fig.~\ref{fig:band_Te}.
The band dispersion displays a ``camel-back'' feature which is clearly
seen in the left panel of Fig.~\ref{fig:band_Te}, with a local minimum
at H (a saddle point in three dimensions), and a maximum about
$0.02\,\,\text{\AA}^{-1}$ close to it.  In the present calculation,
the depth of the minimum with respect to the maximum is 1.3~meV.  The
experimental value is 1.1~meV~\cite{Te_infrared_Doi1970}, the
calculated value with an LAPW code and also a modified Becke-Johnson
functional is 1.7~meV~\cite{Te_LAPW_Snyder2014}, and the calculated
value with the VASP code and the HSE06 functional is
0.8~meV~\cite{Tsirkin_Te_2018}.  These are very small energies, and
while the order of magnitude is consistent, there is some variation
associated with different computational methods and functionals.

\begin{figure} 
\centering
\includegraphics[width=0.6\columnwidth]{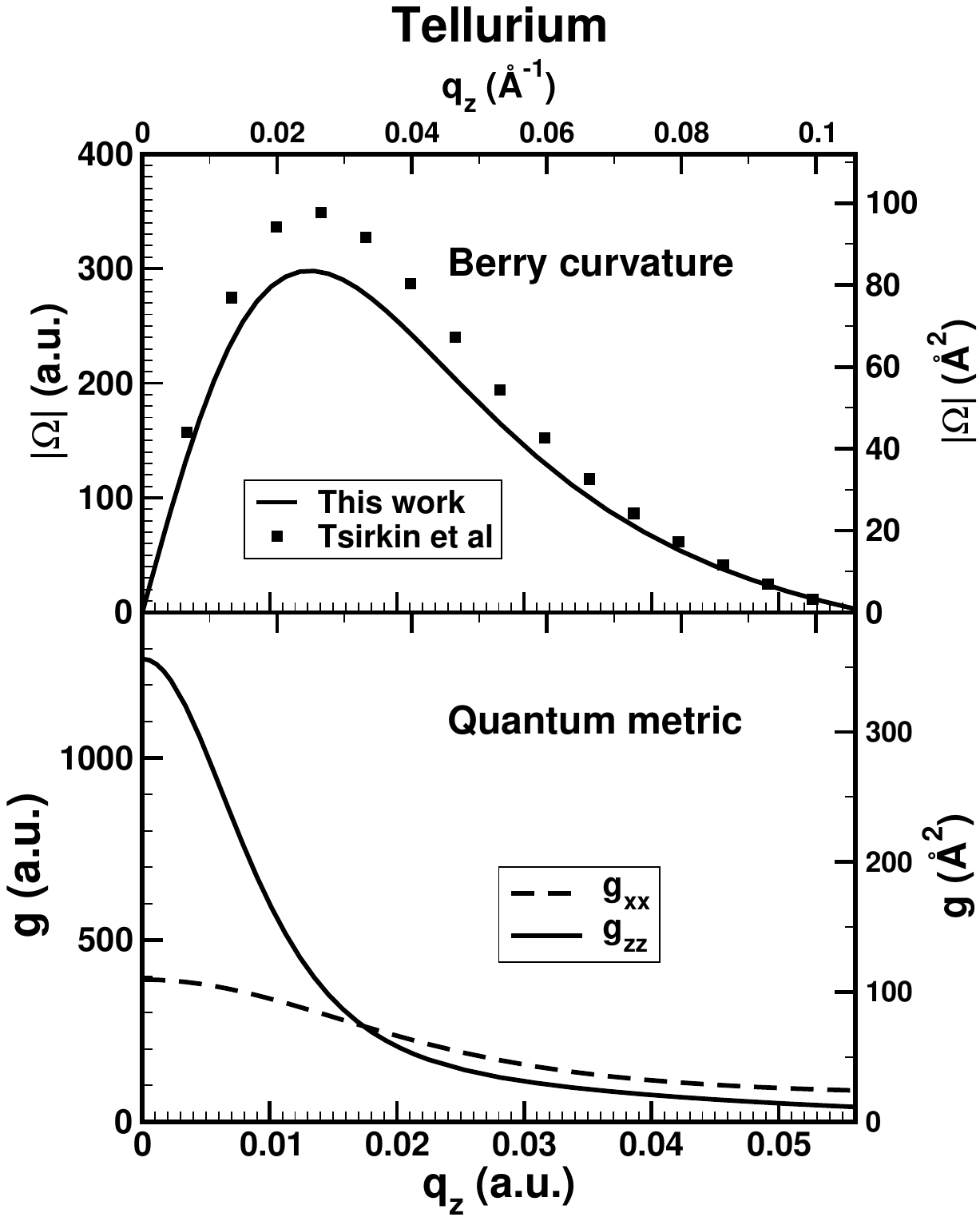}
\caption{The Berry curvature (top panel) and quantum metric (bottom
  panel) of the upper valence band of Te are shown along the H--K line
  as a function of distance from the H point.  In the top panel the
  solid line is the Berry curvature from the present calculation with
  a modified Becke-Johnson functional, and the squares are from a
  calculation by Tsirkin {\it et al.}~\cite{Tsirkin_Te_2018} with the
  HSE06 hybrid functional.  The bottom panel shows the quantum metric
  from the present calculation; the longitudinal component is
  represented by the solid line, and the transverse one by the dashed
  line.  The left and bottom scales are in atomic units, while the top
  and right scales are in SI units.}
\label{fig:berry_metric_Te}
\end{figure}

The Berry curvature near the VBM is shown in the top panel of
Fig.~\ref{fig:berry_metric_Te} along the H--K line, as a function of
the distance from the H point.  Its sign depends on the handedness of
the structure, and which of the two translationally-inequivalent H
points is chosen.  The solid line is the present calculation, and the
squares are the results from a previous calculation with the HSE06
functional~\cite{Tsirkin_Te_2018}.  While the curves are similar, the
values calculated with the present method are slightly smaller.  We
speculate that the difference could be due to the use of
different families of functionals: meta-GGA in the present
work, hybrid functional in Ref.~\cite{Tsirkin_Te_2018}.  The curvature
vanishes at H by symmetry.

The quantum metric near the VBM is shown in the bottom panel of
Fig.~\ref{fig:berry_metric_Te}.  At H the longitudinal component is
significantly larger than the transverse component, but it decays much
faster and becomes smaller outside the camel-back region.

\begin{figure} 
\centering
\includegraphics[width=0.6\columnwidth]{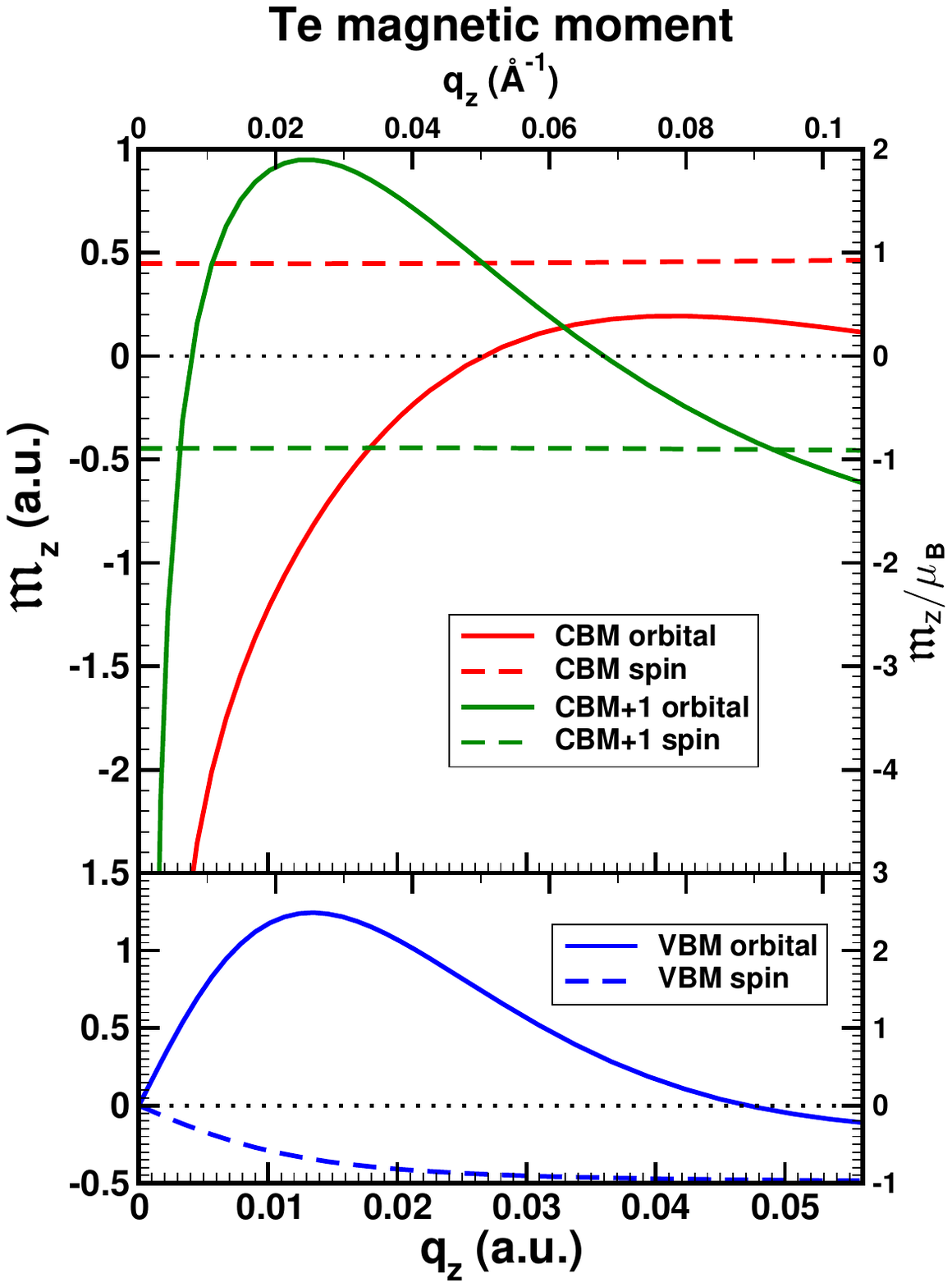}
\caption{The orbital and spin magnetic moments of the upper valence
  band (VBM) and of two lowest conduction bands (CBM and CBM+1) of Te
  are shown along the H--K line as a function of distance from the H
  point.  The solid and dashed lines are respectively the orbital and
  spin magnetic moments.}
\label{fig:mag_Te}
\end{figure}

The orbital and spin magnetic moments near the VBM are shown in
the bottom panel of Fig.~\ref{fig:mag_Te}.  At the H point the upper
valence band is nondegenerate, and both
moments are zero by symmetry,
just like the Berry curvature in Fig.~\ref{fig:berry_metric_Te}.  As
the distance from H increases along the H--K line, the spin magnetic
moment saturates to the Bohr magneton, while the orbital magnetic
moment, which has the opposite sign, first increases and then
decreases.  In both cases the variations with wavevector are on the
scale of the camel-back
feature in the band energies.  The same quantities are shown in the
top panel for the two bands near the conduction band minimum (CBM).
Those bands are degenerate at H, where they have equal and opposite
nearly-saturated spin magnetizations.  The variation of the spin
magnetization with $k$ vector is negligible in the range of the
figure.  Instead, the orbital magnetic moment strongly diverges near
H, as the two-band model from the previous section would predict in
the limit of a vanishing gap.  Near H the orbital moment of the lowest
band has the opposite sign of the spin polarization, but away from H
it decays and changes sign.  In the second-lowest conduction band, the
orbital moment decays and changes sign much faster.  For the upper
valence band, the results are similar and consistent with a previous
calculation~\cite{Tsirkin_Te_2018}.

The signs of $\Omega_z$ and ${\mathfrak m}_z$ change depending on the
handedness of the crystal, on which inequivalent H point one
considers, and, in Fig.~\ref{fig:berry_metric_Te}, on whether $q_z$ is
on the $+z$ or $-z$ direction.  The relative signs of the orbital and
spin magnetic moments and Berry curvature between the different bands
do not change, there is only a global sign flip.  As mentioned
previously, the plotted results are for the left-handed structure.
With the usual $2\pi/3$ angle between the in-plane primitive lattice
vectors, and with the first Te atom at $u=(0.269,0,0)$ in lattice
coordinates, the H point of the figure is at $(1/3,1/3,1/2)$ in
reciprocal lattice coordinates, at a $\pi/3$ angle in the basal plane
with respect to the projection of the atomic positions.  The $q_z$ is
in the negative direction as indicated in the inset of
Fig.~\ref{fig:band_Te},
${\bf k} = {\bf b}_1/3 + {\bf b}_2/3 + (1/2-q_z){\bf b}_3$, and
$\Omega_z$ along that line has negative values.

\begin{figure} 
\centering
\includegraphics[width=0.8\columnwidth]{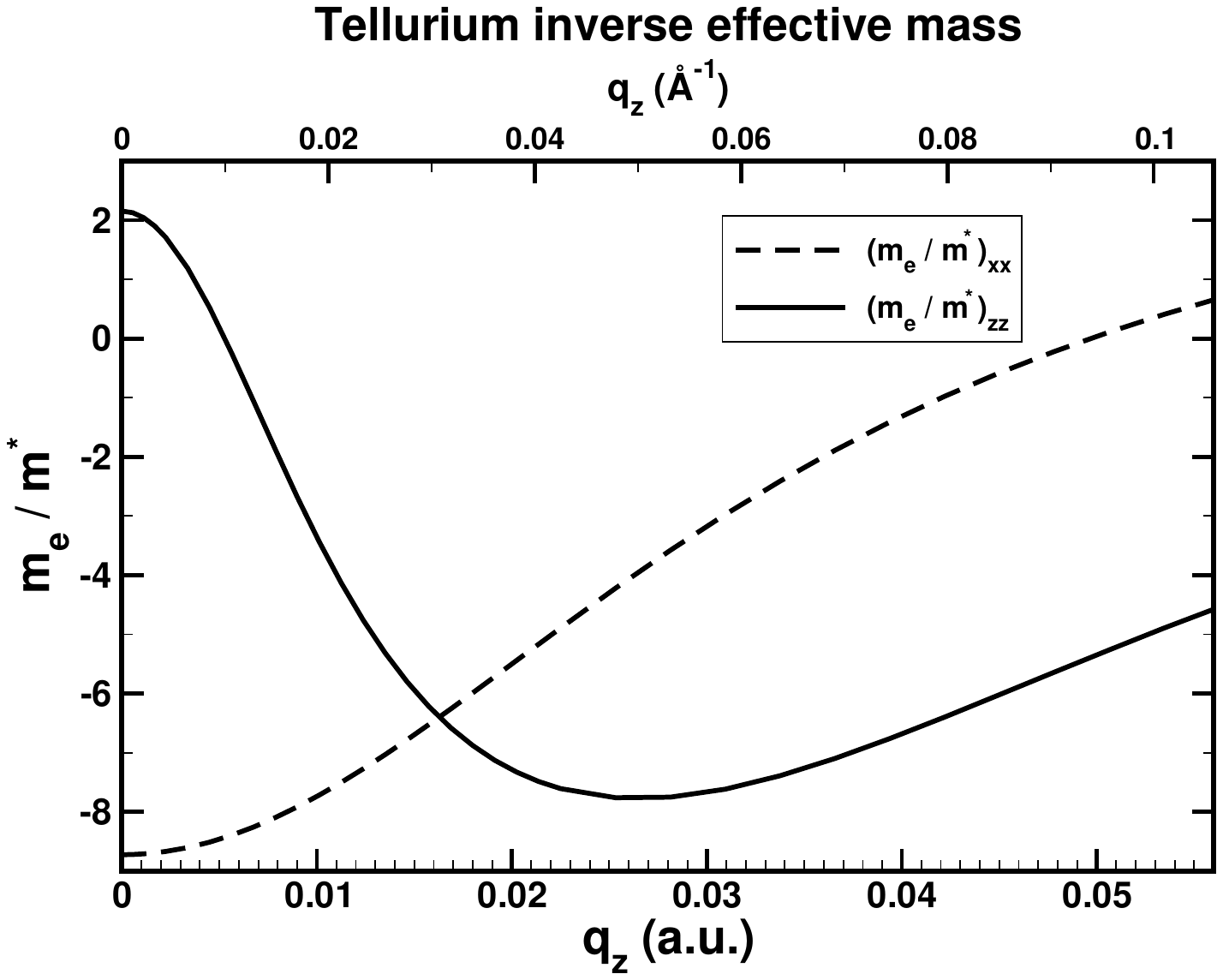}
\caption{The inverse effective mass of the top of the valence band of
  Te is shown along the H--K line as a function of distance from the H
  point.  The solid line is the longitudinal component, and the dashed
  line is the transverse component.  Consistent with the
  ``camel-back'' feature of that band, the longitudinal mass changes
  from positive, electron-like, to negative, hole-like.  The
  transverse mass remains hole-like in almost the entire range of the
  plot.}
\label{fig:mass_Te}
\end{figure}

The components of the inverse effective mass of the upper valence band
of Te are shown in Fig.~\ref{fig:mass_Te} along the H--K line. The
longitudinal mass at H is electron-like while the transverse is
hole-like, as expected for a saddle point.
Along that line, the longitudinal effective mass changes sign going
over the ``hump'' of the camel-back,
and at the top of the valence
band it is hole-like.  The maximum is $0.0185\,\,\text{\AA}^{-1}$ away
from H, where the longitudinal mass is $0.311\,\,m_\text{e}$ and the
transverse mass is $0.129\,\,m_\text{e}$.  The corresponding values
from the LAPW modified Becke-Johnson calculation are
$0.251\,\,m_\text{e}$ and
$0.098\,\,m_\text{e}$~\cite{Te_LAPW_Snyder2014}, while the
experimental values are $0.220\,\,m_\text{e}$ and
$0.108\,\,m_\text{e}$~\cite{Te_mass_1973}.


\section{Conclusions}
\label{sec:conclusions}

The numerically precise calculation of the basic quantum-geometric
properties of crystals, namely Berry curvature, quantum metric,
orbital magnetic moment and effective mass, was successfully
implemented in a first-principles pseudopotential plane-wave code.
The adopted procedure is as follows.  For a given effective potential
and wavevector ${\bf k}$, the first derivatives of the cell-periodic
Bloch wavefunctions $u_{d n {\bf k}} ({\bf r})$ with respect to
wavevector ${\bf k}$ are obtained from perturbation theory.  This is
achieved by solving iteratively a Sternheimer equation, taking into
account energy-level degeneracies and using a stable algorithm.  Once
those wavefunction derivatives are obtained, the four
quantum-geometric quantities are readily determined as the real and
imaginary parts of two complex objects: the quantum-geometric tensor
$T^{\alpha\beta}_{dd'}$ of Eq.~\eqref{eq:T} (quantum metric and the
Berry curvature), and the closely-related mass-moment tensor
$\Gamma^{\alpha\beta}_{dd'}$ of Eq.~\eqref{eq:Gamma} (interband
inverse effective mass and orbital magnetic moment).  The intraband
part contribution to the inverse effective mass given by matrix
elements of the operator $H^{\alpha\beta}_{\bf k}$, is then
included according to
Eq.~\eqref{eq:invmass}.

Effective masses can also be calculated by finite differences.  The
excellent agreement between perturbation theory and finite-differences
values for the effective masses of Si and GaAs shows that they are
precisely calculated with our implementation. This in turn implies a
precise determination of the underlying wavefunction derivatives, and
therefore, the numerical precision of the other quantum-geometric
quantities as well.

In both Si and GaAs, the top of the valence band has nontrivial
degeneracies.  As a result, the $T$ and $\Gamma$ tensors have a total
of four indices: two for the spatial dimensions ($\alpha\beta$), and
two for the level degeneracy ($dd^\prime$).  The calculation of the
direction-dependent transport equivalent effective masses illustrates
how to extract physical properties from those complex objects, taking
into account the subtleties of perturbation theory for degenerate
states.

Calculations of quantum-geometric quantities for gapped graphene were
performed by adding a bespoke artificial potential to first-principles
graphene.  As the two bands near the opened gap are reasonably
decoupled from other bands, in particular once mirror-symmetry
selection rules are accounted for, the first-principles results can be
compared to a well-known two-band low-energy model.  With appropriate
parameters, that model closely reproduced the calculated DFT values,
apart from some small deviations that were discussed.  Such
detailed comparison between low-energy models and direct {\it ab
  initio} calculations has not, to our knowledge, been reported in the
literature.

Trigonal tellurium, one of the simplest crystals with interesting
quantum-geometric properties, was chosen as the final example.  The
properties were calculated in the neighborhood of the valence-band and
conduction-band edges.  For the camel-back region of the valence-band
maximum, the calculated orbital magnetic moment was similar to the
results of a previous calculation.  In the low-lying conduction bands,
the orbital moment near the H point was found to have the opposite
sign from that of the upper valence band.  This could have interesting
implications for nonlinear magneto-transport properties such as
electrical magnetochiral anisotropy~\cite{suarez-natmater25}.

In view of current interest in the quantum geometry of electron states
in crystals and its physical manifestations, we believe that the
formalism and implementation presented in this work could find
multiple applications.

\section*{Acknowledgments}

Work by J.~L.~M.\ and C.~L.~R.\ was supported by grant
EXPL-FIS-MAC-1334-2021 from the Portuguese Science and Technology
Foundation (FCT).  They wish to acknowledge FCT for funding the
Research Unit INESC MN (UID/05367/2020) through Plurianual, Base and
Programático financing.  Work by I.~S.\ was supported by Grant No.
PID2021-129035NB-I00 funded by MCIN/AEI/10.13039/501100011033 and by
ERDF/EU.  The authors thank C.\ J.\ Augusto of Quantum Semiconductor
LLC for useful discussions throughout this work, and for a careful
reading of the manuscript.

\section*{References}

\bibliography{berry,bib}{}

\begin{thebibliography}{10}
\providecommand{\url}[1]{\texttt{#1}}
\providecommand{\urlprefix}{URL }
\expandafter\ifx\csname urlstyle\endcsname\relax
  \providecommand{\doi}[1]{doi:\discretionary{}{}{}#1}\else
  \providecommand{\doi}{doi:\discretionary{}{}{}\begingroup
  \urlstyle{rm}\Url}\fi
\providecommand{\eprint}[2][]{\url{#2}}

\bibitem{xiao-rmp10}
D.~Xiao, M.-C. Chang and Q.~Niu,
\newblock \emph{Berry phase effects on electronic properties},
\newblock Rev. Mod. Phys. \textbf{82}, 1959 (2010),
\newblock \doi{10.1103/RevModPhys.82.1959}.

\bibitem{vanderbilt-book18}
D.~Vanderbilt,
\newblock \emph{{Berry Phases in Electronic Structure Theory: Electric
  Polarization, Orbital Magnetization and Topological Insulators}},
\newblock Cambridge University Press,
\newblock \doi{10.1017/9781316662205} (2018).

\bibitem{yu2025quantumgeometryquantummaterials}
J.~Yu, B.~A. Bernevig, R.~Queiroz, E.~Rossi, P.~Törmä and B.-J. Yang,
\newblock \emph{Quantum geometry in quantum materials},
\newblock \doi{10.48550/arXiv.2501.00098} (2025), \eprint{2501.00098}.

\bibitem{Gao2025quantumgeometry}
A.~Gao, N.~Nagaosa, N.~N and S.-Y. Xu,
\newblock \emph{Quantum geometry phenomena in condensed matter systems},
\newblock \doi{10.48550/arXiv.2508.00469} (2025), \eprint{2508.00469}.

\bibitem{yao-prl04}
Y.~Yao, L.~Kleinman, A.~H. MacDonald, J.~Sinova, T.~Jungwirth, D.-S. Wang,
  E.~Wang and Q.~Niu,
\newblock \emph{{First Principles Calculation of Anomalous Hall Conductivity in
  Ferromagnetic bcc Fe}},
\newblock Phys. Rev. Lett. \textbf{92}, 037204 (2004),
\newblock \doi{10.1103/PhysRevLett.92.037204}.

\bibitem{wang-prb06}
X.~Wang, J.~R. Yates, I.~Souza and D.~Vanderbilt,
\newblock \emph{{{\it Ab initio} calculation of the anomalous Hall conductivity
  by Wannier interpolation}},
\newblock Phys. Rev. B \textbf{74}, 195118 (2006),
\newblock \doi{10.1103/PhysRevB.74.195118}.

\bibitem{lopez-prb12}
M.~G. Lopez, D.~Vanderbilt, T.~Thonhauser and I.~Souza,
\newblock \emph{Wannier-based calculation of the orbital magnetization in
  crystals},
\newblock Phys. Rev. B \textbf{85}, 014435 (2012),
\newblock \doi{10.1103/PhysRevB.85.014435}.

\bibitem{marrazzo-rmp24}
A.~Marrazzo, S.~Beck, E.~R. Margine, N.~Marzari, A.~A. Mostofi, J.~Qiao,
  I.~Souza, S.~S. Tsirkin, J.~R. Yates and G.~Pizzi,
\newblock \emph{Wannier-function software ecosystem for materials simulations},
\newblock Rev. Mod. Phys. \textbf{96}, 045008 (2024),
\newblock \doi{10.1103/RevModPhys.96.045008}.

\bibitem{urru-prb25}
A.~Urru, I.~Souza, O.~P. Oca\~na, S.~S. Tsirkin and D.~Vanderbilt,
\newblock \emph{{Optical spatial dispersion via Wannier interpolation}},
\newblock Phys. Rev. B \textbf{112}, 045201 (2025),
\newblock \doi{10.1103/56cw-5h19}.

\bibitem{Gonze_mass_2016}
J.~Laflamme~Janssen, Y.~Gillet, S.~Ponc\'e, A.~Martin, M.~Torrent and X.~Gonze,
\newblock \emph{Precise effective masses from density functional perturbation
  theory},
\newblock Phys. Rev. B \textbf{93}, 205147 (2016),
\newblock \doi{10.1103/PhysRevB.93.205147}.

\bibitem{BZ_integration_Pickard_1999}
C.~J. Pickard and M.~C. Payne,
\newblock \emph{{Extrapolative approaches to Brillouin-zone integration}},
\newblock Phys. Rev. B \textbf{59}, 4685 (1999),
\newblock \doi{10.1103/PhysRevB.59.4685}.

\bibitem{kdotp_nitrides_JRLeite_2001}
L.~E. Ramos, L.~K. Teles, L.~M.~R. Scolfaro, J.~L.~P. Castineira, A.~L. Rosa
  and J.~R. Leite,
\newblock \emph{{Structural, electronic, and effective-mass properties of
  silicon and zinc-blende group-III nitride semiconductor compounds}},
\newblock Phys. Rev. B \textbf{63}, 165210 (2001),
\newblock \doi{10.1103/PhysRevB.63.165210}.

\bibitem{Blaha_mass_2021}
O.~Rubel, F.~Tran, X.~Rocquefelte and P.~Blaha,
\newblock \emph{Perturbation approach to ab initio effective mass
  calculations},
\newblock Comput. Phys. Commun. \textbf{261}, 107648 (2021),
\newblock \doi{10.1016/j.cpc.2020.107648}.

\bibitem{marzari-prb97}
N.~Marzari and D.~Vanderbilt,
\newblock \emph{{Maximally localized generalized Wannier functions for
  composite energy bands}},
\newblock Phys. Rev. B \textbf{56}, 12847 (1997),
\newblock \doi{10.1103/PhysRevB.56.12847}.

\bibitem{ceresoli-prb06}
D.~Ceresoli, T.~Thonhauser, D.~Vanderbilt and R.~Resta,
\newblock \emph{{Orbital magnetization in crystalline solids: Multi-band
  insulators, Chern insulators, and metals}},
\newblock Phys. Rev. B \textbf{74}, 024408 (2006),
\newblock \doi{10.1103/PhysRevB.74.024408}.

\bibitem{cpw2000}
J.~L. Martins \emph{et~al.},
\newblock \emph{cpw2000},
\newblock \doi{10.5281/zenodo.15765091} (2025).

\bibitem{Te_LAPW_Snyder2014}
H.~Peng, N.~Kioussis and G.~J. Snyder,
\newblock \emph{Elemental tellurium as a chiral $p$-type thermoelectric
  material},
\newblock Phys. Rev. B \textbf{89}, 195206 (2014),
\newblock \doi{10.1103/PhysRevB.89.195206}.

\bibitem{lin-natcomms16}
S.~Lin, W.~Li, Z.~Chen, J.~Shen, B.~Ge and Y.~Pei,
\newblock \emph{Tellurium as a high-performance elemental thermoelectric},
\newblock Nat. Commun. \textbf{7}, 10287 (2016),
\newblock \doi{10.1038/ncomms10287}.

\bibitem{Tsirkin_Te_2018}
S.~S. Tsirkin, P.~A. Puente and I.~Souza,
\newblock \emph{Gyrotropic effects in trigonal tellurium studied from first
  principles},
\newblock Phys. Rev. B \textbf{97}, 035158 (2018),
\newblock \doi{10.1103/PhysRevB.97.035158}.

\bibitem{sahin-prb18}
C.~\ifmmode~\mbox{\c{S}}\else \c{S}\fi{}ahin, J.~Rou, J.~Ma and D.~A. Pesin,
\newblock \emph{{Pancharatnam-Berry phase and kinetic magnetoelectric effect in
  trigonal tellurium}},
\newblock Phys. Rev. B \textbf{97}, 205206 (2018),
\newblock \doi{10.1103/PhysRevB.97.205206}.

\bibitem{furukawa-prb21}
T.~Furukawa, Y.~Watanabe, N.~Ogasawara, K.~Kobayashi and T.~Itou,
\newblock \emph{Current-induced magnetization caused by crystal chirality in
  nonmagnetic elemental tellurium},
\newblock Phys. Rev. Res. \textbf{3}, 023111 (2021),
\newblock \doi{10.1103/PhysRevResearch.3.023111}.

\bibitem{nakazawa-prm24}
K.~Nakazawa, T.~Yamaguchi and A.~Yamakage,
\newblock \emph{Nonlinear charge transport properties in chiral tellurium},
\newblock Phys. Rev. Mater. \textbf{8}, L091601 (2024),
\newblock \doi{10.1103/PhysRevMaterials.8.L091601}.

\bibitem{pan-prb25}
M.~Pan, H.~Zeng, E.~Wang and H.~Huang,
\newblock \emph{{Intrinsic orbital origin for the chirality-dependent nonlinear
  planar Hall effect of topological nodal fermions in chiral crystals}},
\newblock Phys. Rev. B \textbf{111}, 075145 (2025),
\newblock \doi{10.1103/PhysRevB.111.075145}.

\bibitem{KohnSham1965}
W.~Kohn and L.~J. Sham,
\newblock \emph{{Self-Consistent Equations Including Exchange and Correlation
  Effects}},
\newblock Phys. Rev. \textbf{140}, A1133 (1965),
\newblock \doi{10.1103/PhysRev.140.A1133}.

\bibitem{HohenbergKohn1964}
P.~Hohenberg and W.~Kohn,
\newblock \emph{{Inhomogeneous Electron Gas}},
\newblock Phys. Rev. \textbf{136}, B864 (1964),
\newblock \doi{10.1103/PhysRev.136.B864}.

\bibitem{martin-book04}
R.~M. Martin,
\newblock \emph{{Electronic Structure: Basic Theory and Practical Methods}},
\newblock Cambridge, 1st edn. (2004).

\bibitem{KleinmanBylander1982}
L.~Kleinman and D.~M. Bylander,
\newblock \emph{{Efficacious Form for Model Pseudopotentials}},
\newblock Phys. Rev. Lett. \textbf{48}, 1425 (1982),
\newblock \doi{10.1103/PhysRevLett.48.1425}.

\bibitem{MIT-OCW_Zwiebach_2018}
B.~Zwiebach,
\newblock \emph{Mastering Quantum Mechanics},
\newblock The MIT Press,
\newblock ISBN 9780262046138 (2022).

\bibitem{CarParrinello1985}
R.~Car and M.~Parrinello,
\newblock \emph{{Unified Approach for Molecular Dynamics and Density-Functional
  Theory}},
\newblock Phys. Rev. Lett. \textbf{55}, 2471 (1985),
\newblock \doi{10.1103/PhysRevLett.55.2471}.

\bibitem{MartinsCohen1988}
J.~L. Martins and M.~L. Cohen,
\newblock \emph{{Diagonalization of large matrices in pseudopotential
  band-structure calculations: Dual-space formalism}},
\newblock Phys. Rev. B \textbf{37}, 6134 (1988),
\newblock \doi{10.1103/PhysRevB.37.6134}.

\bibitem{ProvostVallee_1980}
J.~P. Provost and G.~Vallee,
\newblock \emph{Riemannian structure on manifolds of quantum states},
\newblock Commun. Math. Phys. \textbf{76}, 289 (1980),
\newblock \doi{10.1007/BF02193559}.

\bibitem{berry1989}
M.~V. Berry,
\newblock \emph{{The Quantum Phase, Five years After}},
\newblock In F.~Wilczek and A.~Shapere, eds., \emph{{Geometric Phases in
  Physics}}. World Scientific, Singapore (1989).

\bibitem{souza-prb08}
I.~Souza and D.~Vanderbilt,
\newblock \emph{Dichroic $f$-sum rule and the orbital magnetization of
  crystals},
\newblock Phys. Rev. B \textbf{77}, 054438 (2008),
\newblock \doi{10.1103/PhysRevB.77.054438}.

\bibitem{chang-jpcm08}
M.-C. Chang and Q.~Niu,
\newblock \emph{{Berry curvature, orbital moment, and effective quantum theory
  of electrons in electromagnetic fields}},
\newblock J. Phys. Condens. Matter \textbf{20}, 193202 (2008),
\newblock \doi{10.1088/0953-8984/20/19/193202}.

\bibitem{Resta_2018}
R.~Resta,
\newblock \emph{Drude weight and superconducting weight},
\newblock J. Phys. Condens. Matter \textbf{30}, 414001 (2018),
\newblock \doi{10.1088/1361-648X/aade19}.

\bibitem{Band_warp_Resca2014}
N.~A. Mecholsky, L.~Resca, I.~L. Pegg and M.~Fornari,
\newblock \emph{Theory of band warping and its effects on thermoelectronic
  transport properties},
\newblock Phys. Rev. B \textbf{89}, 155131 (2014),
\newblock \doi{10.1103/PhysRevB.89.155131}.

\bibitem{kdotp_masses_Pickard2000}
C.~J. Pickard and M.~C. Payne,
\newblock \emph{Second-order $\mathbf{k}\ensuremath{\cdot}\mathbf{p}$
  perturbation theory with vanderbilt pseudopotentials and plane waves},
\newblock Phys. Rev. B \textbf{62}, 4383 (2000),
\newblock \doi{10.1103/PhysRevB.62.4383}.

\bibitem{PerdewWang1992}
J.~P. Perdew and Y.~Wang,
\newblock \emph{Accurate and simple analytic representation of the electron-gas
  correlation energy},
\newblock Phys. Rev. B \textbf{45}, 13244 (1992),
\newblock \doi{10.1103/PhysRevB.45.13244}.

\bibitem{TroullierMartins1991-I}
N.~Troullier and J.~L. Martins,
\newblock \emph{{Efficient pseudopotentials for plane-wave calculations}},
\newblock Phys. Rev. B \textbf{43}, 1993 (1991),
\newblock \doi{10.1103/PhysRevB.43.1993}.

\bibitem{atom}
S.~Froyen, N.~Troullier, J.~L. Martins \emph{et~al.},
\newblock \emph{Pseudopotential generation code},
\newblock \doi{10.5281/zenodo.15765105} (2025).

\bibitem{poly_interp}
J.~L. Martins and C.~L. Reis,
\newblock \emph{Lagrange polynomial interpolation with derivatives and error
  indication},
\newblock \doi{10.5281/zenodo.15425573} (2025).

\bibitem{LuttingerKohn}
J.~Luttinger and W.~Kohn,
\newblock \emph{Motion of electrons and holes in perturbed periodic fields},
\newblock Phys. Rev. \textbf{97}, 869 (1955),
\newblock \doi{10.1103/PhysRev.97.869}.

\bibitem{Dresselhaus_ZB}
G.~Dresselhaus,
\newblock \emph{Spin-orbit coupling effects in zinc blende structures},
\newblock Phys. Rev. \textbf{100}, 580 (1955),
\newblock \doi{10.1103/PhysRev.100.580}.

\bibitem{TranBlaha2009}
F.~Tran and P.~Blaha,
\newblock \emph{Accurate band gaps of semiconductors and insulators with a
  semilocal exchange-correlation potential},
\newblock Phys. Rev. Lett. \textbf{102}, 226401 (2009),
\newblock \doi{10.1103/PhysRevLett.102.226401}.

\bibitem{graphene_epitaxial_2007}
S.~Y. Zhou, G.-H. Gweon, A.~V. Fedorov, P.~N. First, W.~A. de~Heer, D.-H. Lee,
  F.~Guinea, A.~H. Castro~Neto and A.~Lanzara,
\newblock \emph{Substrate-induced bandgap opening in epitaxial graphene},
\newblock Nat. Mater. \textbf{6}, 770 (2007),
\newblock \doi{10.1038/nmat2003}.

\bibitem{review_graphene_Neto_Peres_2009}
A.~H. Castro~Neto, F.~Guinea, N.~M.~R. Peres, K.~S. Novoselov and A.~K. Geim,
\newblock \emph{The electronic properties of graphene},
\newblock Rev. Mod. Phys. \textbf{81}, 109 (2009),
\newblock \doi{10.1103/RevModPhys.81.109}.

\bibitem{graphite_TB_Wallace_1947}
P.~R. Wallace,
\newblock \emph{The band theory of graphite},
\newblock Phys. Rev. \textbf{71}, 622 (1947),
\newblock \doi{10.1103/PhysRev.71.622}.

\bibitem{graphene_SO_Kane_Mele_2005}
C.~L. Kane and E.~J. Mele,
\newblock \emph{{Quantum Spin Hall Effect in Graphene}},
\newblock Phys. Rev. Lett. \textbf{95}, 226801 (2005),
\newblock \doi{10.1103/PhysRevLett.95.226801}.

\bibitem{pozo-prb20}
O.~Pozo and F.~de~Juan,
\newblock \emph{{Computing observables without eigenstates: Applications to
  Bloch Hamiltonians}},
\newblock Phys. Rev. B \textbf{102}, 115138 (2020),
\newblock \doi{10.1103/PhysRevB.102.115138}.

\bibitem{graf-prb21}
A.~Graf and F.~Pi\'echon,
\newblock \emph{{Berry curvature and quantum metric in $N$-band systems: An
  eigenprojector approach}},
\newblock Phys. Rev. B \textbf{104}, 085114 (2021),
\newblock \doi{10.1103/PhysRevB.104.085114}.

\bibitem{XiaoYaoNiu2007}
D.~Xiao, W.~Yao and Q.~Niu,
\newblock \emph{Valley-contrasting physics in graphene: Magnetic moment and
  topological transport},
\newblock Phys. Rev. Lett. \textbf{99}, 236809 (2007),
\newblock \doi{10.1103/PhysRevLett.99.236809}.

\bibitem{Schabel_graphite}
M.~C. Schabel and J.~L. Martins,
\newblock \emph{Energetics of interplanar binding in graphite},
\newblock Phys. Rev. B \textbf{46}, 7185 (1992),
\newblock \doi{10.1103/PhysRevB.46.7185}.

\bibitem{roy-prb14}
R.~Roy,
\newblock \emph{Band geometry of fractional topological insulators},
\newblock Phys. Rev. B \textbf{90}, 165139 (2014),
\newblock \doi{10.1103/PhysRevB.90.165139}.

\bibitem{Mera2021}
T.~Ozawa and B.~Mera,
\newblock \emph{{Relations between topology and the quantum metric for Chern
  insulators}},
\newblock Phys. Rev. B \textbf{104}, 045103 (2021),
\newblock \doi{10.1103/PhysRevB.104.045103}.

\bibitem{Te_Asendorf1957}
R.~H. Asendorf,
\newblock \emph{Space group of tellurium and selenium},
\newblock J. Chem. Phys. \textbf{27}, 11 (1957),
\newblock \doi{10.1063/1.1743647}.

\bibitem{Te_GW_2015}
M.~Hirayama, R.~Okugawa, S.~Ishibashi, S.~Murakami and T.~Miyake,
\newblock \emph{Weyl node and spin texture in trigonal tellurium and selenium},
\newblock Phys. Rev. Lett. \textbf{114}, 206401 (2015),
\newblock \doi{10.1103/PhysRevLett.114.206401}.

\bibitem{Te_gap_eremets1977}
V.~B. Anzin, M.~I. Eremets, Y.~V. Kosichkin, A.~I. Nadezhdinskii and A.~M.
  Shirokov,
\newblock \emph{Measurement of the energy gap in tellurium under pressure},
\newblock Phys. Status Solidi (a) \textbf{42}, 385 (1977),
\newblock \doi{10.1002/pssa.2210420143}.

\bibitem{Te_infrared_Doi1970}
T.~Doi, K.~Nakao and H.~Kamimura,
\newblock \emph{The valence band structure of tellurium. ii. the infrared
  absorption},
\newblock J. Phys. Soc. Jpn. \textbf{28}, 822 (1970),
\newblock \doi{10.1143/JPSJ.28.822}.

\bibitem{Te_mass_1973}
Y.~Couder, M.~Hulin and H.~Thom\'e,
\newblock \emph{Cyclotron resonance in tellurium},
\newblock Phys. Rev. B \textbf{7}, 4373 (1973),
\newblock \doi{10.1103/PhysRevB.7.4373}.

\bibitem{suarez-natmater25}
M.~Suárez-Rodríguez, F.~{de Juan}, I.~Souza, M.~Gobbi, F.~Casanova and L.~E.
  Hueso,
\newblock \emph{Nonlinear transport in non-centrosymmetric systems},
\newblock Nat. Mater. \textbf{24}, 1005 (2025),
\newblock \doi{10.1038/s41563-025-02261-3}.

\end{thebibliography}

\end{document}